%% file: ex_article_v1.tex
\documentclass[review,onefignum,onetabnum]{siamonline250211}

\usepackage{amsmath, xcolor, tikz, subcaption, amssymb} % Required for inserting images
\usetikzlibrary{positioning}
\definecolor{processblue}{cmyk}{0.96,0,0,0}
\usepackage[margin=1in]{geometry}

\input{ex_shared}

\ifpdf
\hypersetup{
  pdftitle={Modeling microbiome dynamics on social networks: how between-host transmission shapes within-host evolution},
  pdfauthor={Longmei Shu and Feng Fu}
}
\fi

\begin{document}

\maketitle

% REQUIRED
\begin{abstract}
A social network perspective on the human microbiome is crucial for understanding how the interplay between within-host microbial dynamics and between-host transmission influences microbial community stability, host metabolism, and population health and nutrition. Here, we provide a mathematical and numerical analysis of human-associated microbial communities in dyadic social ties and within a small network. By utilizing a generalized Lotka-Volterra model, we simulate within-host dynamics while incorporating inter-host microbial sharing driven by social interactions. Furthermore, the study accounts for individual variability in microbial behavior across different hosts. Our work identifies the conditions under which microbial sharing can have long-lasting effects, offering insights into how social networks shape the human microbiome. 
\end{abstract}

% REQUIRED

% REQUIRED

\section{Introduction}

Human-associated microbial communities play key roles in host health, metabolism, and immunity, and resistance against pathogens. Large-scale surveys have revealed that such communities are highly diverse and personalized with substantial variation in both taxonomic composition and relative abundance across individuals, body sites, geographic populations, and lifestyles \cite{hmp2012,yatsunenko2012,falony2016,zhernakova2016}. Despite this interindividual heterogeneity, an individual's microbial community can be remarkably persistent over long periods of time, with many strains persisting for years \cite{faith2013}. At the same time, microbial communities are dynamic and can respond rapidly to environmental and behavioral changes, such as variations in diet and exposure to antibiotics \cite{dethlefsen2011,david2014}. Thus, the human microbiome can be viewed as a dynamic ecological system with strong endogenous stability and the capacity for large and sometimes persistent changes.

The ecological nature of the microbiome has motivated the development of microbiome-based interventions. Fecal microbiota transplantation (FMT) can restore disrupted microbial communities and have been shown to have substantial efficacy in treating recurrent \textit{Clostridioides difficile} infection \cite{vanNood2013}. Strain-resolved studies of FMT further show that microbes transferred from one individual can successfully engraft and persist within another host, albeit dependent strongly on the ecological properties of both donor and recipient communities \cite{smillie2018}. These observations show that the microbiome of an individual is not a closed ecological system: microbial immigration from other hosts can alter community composition and have long-lasting effects.

Outside of deliberate microbiome transplants, interpersonal contact provides a natural way for microbial communities to become coupled. Cohabiting family members have been shown to share more microbial taxa than unrelated individuals, with frequent physical contact providing opportunities for microbial exchange \cite{song2013}. Evidence from non-human primates provide an especially clear connection between social interactions and gut microbial composition. Social-network relationships and rates of grooming interactions predict microbiome similarity in wild baboons, even after controlling for diet, kinship, and shared environment \cite{tung2015}. Similarly, social interactions among wild chimpanzees are associated with increased gut microbial diversity within individuals, and higher similarity across individuals \cite{moeller2016}. These findings suggest that social organization creates an ecological environment beyond the boundaries of an individual host.

Increasingly fine-grained human metagenomic data provide further evidence of interpersonal microbial sharing. Strain-level analyses in Fijian communities have identified patterns of microbial sharing along family and social relationships, with particularly strong signals for household members and spouses \cite{brito2019}. More recently, large-scale strain-resolved analyses have documented extensive person-to-person sharing of gut and oral microbes, and showed strong associations between microbial sharing, cohabitation, and duration of interpersonal association \cite{vallescolomer2023}. Together, these empirical findings support a social-network perspective on the human microbiome~\cite{beghini25}: each host contains a complex microbial ecosystem, while social interactions create ways for microbes to move between these ecosystems.

Understanding such socially coupled microbial communities requires considering processes occurring at two interconnected scales. Within a given host, microbes compete, facilitate, inhibit, and modify one another's ecological environments. Both competition and cooperation among microbes can strongly affect microbiome composition and host outcomes \cite{coyte19,figueiredo20}. At the between-host scale, microbial dispersal through social interactions couples otherwise distinct host ecosystems. The resulting dynamics depend not only on the rate at which microbes are exchanged, but also whether the transmitted microbes can establish themselves within the ecological environment of the recipient host.

Generalized Lotka--Volterra (gLV) models provide a classical and tractable framework for describing within-host ecological dynamics. They have been widely used to characterize microbial competition, interaction networks, community stability, and responses to perturbations \cite{coyte2015,stein2013}. Methods based on gLV dynamics have also been developed to infer microbial interactions from longitudinal microbiome time series \cite{bucci2016}. Experimental studies of constructed microbial communities show that relatively simple ecological interaction models capture important features of multispecies microbial dynamics and competition \cite{venturelli2018,wang241}. Related modeling and experimental work has shown how interactions among community members determines colonization resistance and microbiome recovery following perturbation \cite{buffie2015}. More broadly, analyses of human microbiome data suggest that microbial communities may exhibit common dynamical principles despite pronounced differences in their taxonomic compositions across individuals \cite{bashan2016}. These findings motivate our use of a gLV framework to represent microbial growth and competition within each host.

Coupling microbial dynamics across hosts naturally connects to metacommunity theory, in which local ecological communities sitting in distinct patches are linked by dispersal. In the context of host-associated microbiomes, individual hosts can be seen as ecological patches, and social interactions as opportunities for microbial exchange. Recent theoretical work by Johnson and Porter formalized this perspective, developing a metacommunity framework for interacting hosts in which microbial exchange occurs over discrete interaction events with interaction frequency and amount of microbial exchange represented separately \cite{johnson25}. Their analysis shows that the interplay between the frequency and strength of host interactions can have substantial effects on convergence of microbial communities across hosts.

In this paper, we take a complementary approach in which microbial sharing occurs continuously, along weighted social ties. Within a host, microbial population dynamics are governed by a generalized Lotka--Volterra competition model. Between hosts, each microbial type is exchanged at a rate proportional to the strength of the social connection and the difference in its abundance between the two hosts. This diffusion-like coupling provides a parsimonious representation of microbial sharing across a social network: stronger social connections yield greater microbial exchange, and differences in microbial abundance generate directional flux between connected hosts. Importantly, microbial sharing does not by itself determine the eventual composition of the recipient microbiome: the fate of an introduced microbial population depends on its interactions with the resident microbial community.

We further allow microbial growth and interaction parameters to vary across hosts, capturing interindividual heterogeneity in the ecological environments encountered by transmitted microbes. This heterogeneity allows social transmission to have qualitatively different consequences among individuals: the same microbial exposure may disappear quickly in one host, and cause a persistent shift in another. Our framework, therefore, links three components that jointly determine socially mediated microbiome dynamics---within-host ecological interactions, between-host microbial sharing, and heterogeneity across hosts.

Using this framework, we ask when microbial sharing through social interactions generates only transient changes, and when it has long-lasting effects on host microbial communities. We first study a two-host, two-strain system in \cref{2-host}, which allows a detailed analysis of the coupled ecological dynamics, and the conditions for persistent effects of microbial sharing. In \cref{num}, we extend our numerical analysis to a two-host, ten-strain system, and then to a small social network consisting of eight hosts, each containing ten microbial strains. We close in \cref{sec:discussions} by discussing the implications and limitations of the model, and its relevance for understanding how social interactions shape the human microbiome.

\section{2-host 2-strain}
\label{2-host}

In this section, we look at the microbial dynamics in 2 hosts, each with 2 strains of microbes. We start by looking at the in-host dynamics and then take inter-host social interactions into consideration.

\subsection{In-host model}

Consider 2 strains of microbes competing in host 1. The system is governed by
\begin{equation}
    \dot x=x(1-x-y),\quad \dot y=\frac{y}{4}(3-2x-4y).
    \label{host1stable}
\end{equation}

\begin{figure}[h]
    \includegraphics[width=\linewidth]{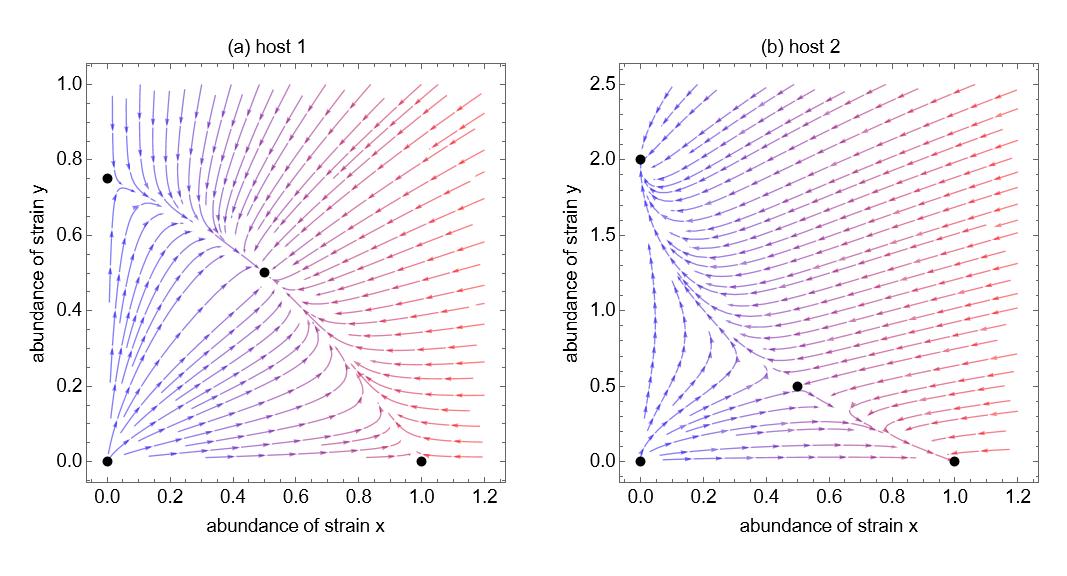}
    \caption{In-host microbial dynamics}
    \label{in-host}
\end{figure}

We can study the dynamics of the system using the phase portrait and the Jacobian matrix. There is a stable coexistence sink at $(0.5,0.5)$, an unstable source at (0,0), and two saddle points (0,0.75), (1,0); see Figure \ref{in-host}(a). 
The Jacobian of the system is $$J=\begin{pmatrix}
    1-2x-y & -x \\ -\frac{y}{2} & \frac{3}{4}-\frac{x}{2} -2y
\end{pmatrix}.$$
When we plug the fixed points into the Jacobian and compute the eigenvalues, we can see that the stability analysis agrees with the phase portrait.

When the parameters change, the dynamics also changes. Consider the same strains of microbes in host 2.
\begin{equation}
    \dot x=x(1-x-y), \quad \dot y=y(0.5-0.75x-0.25y).
    \label{host2}
\end{equation}
There is an unstable co-existence saddle point at $(0.5,0.5)$, an unstable source at (0,0), and two stable sinks (0,2), (1,0); see Figure \ref{in-host}(b). Again we can compute the eigenvalues of the Jacobian $$J=\begin{pmatrix}
    1-2x-y & -x \\ -0.75y & 0.5-0.75x-0.5y
\end{pmatrix}$$ at each fixed point and verify the stability analysis.

\begin {figure}[h]
\centering
\begin{tikzpicture}[-latex, auto, node distance = 0.3 cm and 0.3 cm, on grid, semithick, state/.style ={circle, top color =white, bottom color = processblue!20, draw, processblue, text=blue, minimum width =0.3 cm}]
\node at (0,1) {Host 1};
\node[state] (A) at (0, 0) {$xy$};
\node[state] (B) at (-1.5,-1.5) {$x$};
\node[state] (C) at (0,-3) {$\emptyset$};
\node[state] (D) at (1.5,-1.5) {$y$};
\node at (5,1) {Host 2};
\node[state] (E) at (3.5,-1.5) {$x$};
\node[state] (F) at (5,0) {$xy$};
\node[state] (G) at (6.5,-1.5) {$y$};
\node[state] (H) at (5,-3) {$\emptyset$};
\path (B) edge (A);
\path (D) edge (A);
\path (C) edge (A) edge (B) edge (D);
\path (H) edge (E) edge (F) edge (G);
\path (F) edge (E) edge (G);
\end{tikzpicture}
\caption{Invasion graphs for in-host microbial dynamics}\label{invasion-inhost-2}
\end{figure}
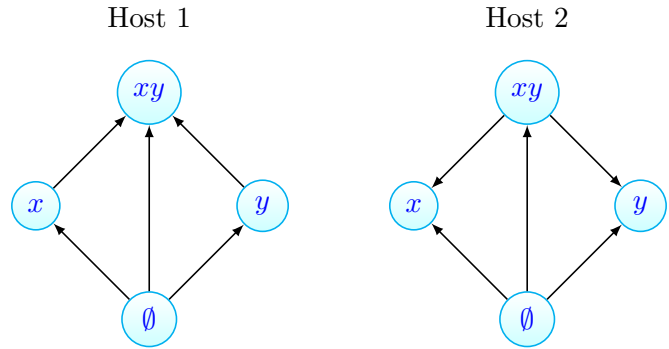

Now we would like to introduce a new tool, the invasion graph \cite{hofbauer22}, to describe the microbial dynamics in the hosts, see Figure \ref{invasion-inhost-2}. Here the $\emptyset$ node represents the fixed point or equilibrium where neither microbe is present. $x$ and $y$ nodes represent the equilibria where only strain $x$ is present and only strain $y$ is present, respectively. And the $xy$ node represents the equilibrium where both strains coexist in the host. The edges are determined by the signs of per-capita growth rates. For example, in host 1, the per-capita growth rates are $$r=(1-x-y,0.75-0.5x-y).$$ At $\emptyset$ the signs of the per-capita growth rates are $g(\emptyset)=(+,+)$. 

There is an edge from node $S$ to node $T$ if 
\begin{enumerate}
    \item[i)] $S\ne T$,
    \item[ii)] $g(S)$ has positive signs on all components in $T\setminus S$,
    \item[iii)] $g(T)$ has negative signs on all components in $S\setminus T$.
\end{enumerate}
The edges follow the dynamics flow of the system. We can see the invasion graphs in Figure \ref{invasion-inhost-2} match the phase portraits in Figure \ref{in-host}. Coexistence is stable in host 1 but unstable in host 2.

\subsection{Inter-host model}
Assume now that host 1 and host 2 have social interactions that update their microbial profile. To clarify, we use $x_1$ to represent the abundance of strain $x$ in host 1 and $x_2$ for the abundance of strain $x$ in host 2. Similarly $y_1$ represents the abundance of strain $y$ in host 1 and $y_2$ represents the abundance of strain $y$ in host 2. Then the new coupled model is
\begin{equation}
    \begin{cases}
        \dot x_1=x_1(1-x_1-y_1)+c_{11}x_1(x_2-x_1),\\
        \dot y_1=\frac{y_1}{4}(3-2x_1-4y_1)+c_{12}y_1(y_2-y_1),\\
        \dot x_2=x_2(1-x_2-y_2)+c_{21}x_2(x_1-x_2),\\
        \dot y_2=y_2(0.5-0.75x_2-0.25y_2)+c_{22}y_2(y_1-y_2).
    \end{cases}
    \label{2-2}
\end{equation}

Here $C=\begin{pmatrix}
    c_{11} & c_{12} \\
    c_{21} & c_{22}
\end{pmatrix}$ depends on the social connection between the two hosts.
For example, let $C=\begin{pmatrix}
    0.1 & 0.1 \\
    0.2 & 0.2
\end{pmatrix}$. Then our system becomes
\begin{equation}
    \begin{cases}
        \dot x_1=x_1(1-1.1x_1-y_1+0.1x_2),\\
        \dot y_1=y_1(0.75-0.5x_1-1.1y_1+0.1y_2),\\
        \dot x_2=x_2(1-1.2x_2-y_2+0.2x_1),\\
        \dot y_2=y_2(0.5-0.75x_2-0.45y_2+0.2y_1).
    \end{cases}
    \label{2-2-asymm}
\end{equation}

We are looking at a 4-dimensional system and the phase portrait will be difficult to visualize. We can still compute the eigenvalues of the Jacobian and will resort to the invasion graph to visualize the dynamics. The Jacobian of the system is 
\begin{align*}
    &J=\\&\tiny\begin{pmatrix}
        1-2.2x_1-y_1+0.1x_2 & -x_1 & 0.1x_1 & 0 \\
        -0.5y_1 & 0.75-0.5x_1-2.2y_1+0.1y_2 & 0 & 0.1y_1\\
        0.2x_2 & 0 & 1-2.4x_2-y_2+0.2x_1 & -x_2 \\
        0 & 0.2y_2 & -0.75y_2 & 0.5-0.75x_2-0.9y_2+0.2y_1
    \end{pmatrix}
\end{align*}
and the per-capita growth rates are $$r=\begin{pmatrix}
    1-1.1x_1-y_1+0.1x_2 \\ 0.75-0.5x_1-1.1y_1+0.1y_2 \\ 1-1.2 x_2-y_2+0.2x_1 \\ 0.5-0.75x_2-0.45y_2+0.2y_1
\end{pmatrix}.$$

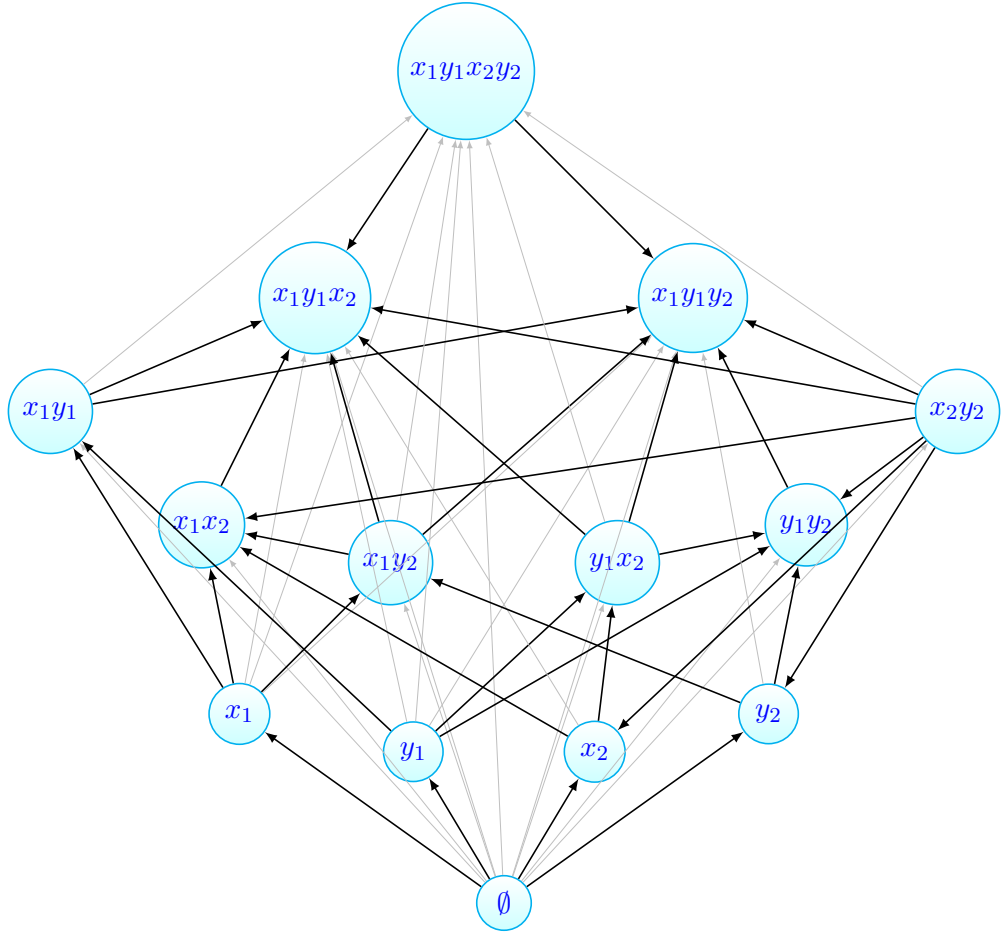
\begin {figure}[h]
\centering
\begin{tikzpicture}[-latex, auto, node distance = 0.3 cm and 0.3 cm, on grid, semithick, state/.style ={circle, top color =white, bottom color = processblue!20, draw, processblue, text=blue, minimum width =0.3 cm}]

\node[state] (A) at (0, 0) {$\emptyset$};
\node[state] (A1) at (-3.5,2.5) {$x_1$};
\node[state] (B1) at (-1.2,2) {$y_1$};
\node[state] (C1) at (1.2,2) {$x_2$};
\node[state] (D1) at (3.5,2.5) {$y_2$};
\node[state] (A2) at (-6,6.5) {$x_1y_1$};
\node[state] (B2) at (-4,5) {$x_1x_2$};
\node[state] (C2) at (-1.5,4.5) {$x_1y_2$};
\node[state] (D2) at (1.5,4.5) {$y_1x_2$};
\node[state] (E2) at (4,5) {$y_1y_2$};
\node[state] (F2) at (6,6.5) {$x_2y_2$};
\node[state] (A3) at (-2.5,8) {$x_1y_1x_2$};
\node[state] (B3) at (2.5,8) {$x_1y_1y_2$};
\node[state] (A4) at (-0.5,11) {$x_1y_1x_2y_2$};
\path (A) edge (A1) edge (B1) edge (C1) edge (D1) edge[lightgray, ultra thin] (A2) edge[lightgray, ultra thin] (B2) edge[lightgray, ultra thin] (C2) edge[lightgray, ultra thin] (D2) edge[lightgray, ultra thin] (E2) edge[lightgray, ultra thin] (F2) edge[lightgray, ultra thin] (A3) edge[lightgray, ultra thin] (B3) edge[lightgray, ultra thin] (A4);
\path (A1) edge (A2) edge (B2) edge (C2) edge[lightgray, ultra thin] (A3) edge[lightgray, ultra thin] (B3) edge[lightgray, ultra thin] (A4);
\path (B1) edge (A2) edge (D2) edge (E2) edge[lightgray, ultra thin] (A3) edge[lightgray, ultra thin] (B3) edge[lightgray, ultra thin] (A4);
\path (C1) edge (B2) edge (D2) edge[lightgray, ultra thin] (A3);
\path (D1) edge (C2) edge (E2) edge[lightgray, ultra thin] (B3);
\path (A2) edge (A3) edge (B3) edge[lightgray, ultra thin] (A4); 
\path (B2) edge (A3);
\path (C2) edge (B2) edge (A3) edge (B3) edge[lightgray, ultra thin] (A4);
\path (D2) edge (E2) edge (A3) edge (B3) edge[lightgray, ultra thin] (A4);
\path (E2) edge (B3);
\path (F2) edge (C1) edge (D1) edge (B2) edge (E2) edge (A3) edge (B3) edge[lightgray, ultra thin] (A4);
\path (A4) edge (A3) edge (B3);
\end{tikzpicture}
\caption{Invasion graph for inter-host microbial dynamics}\label{invasion-inter-2}
\end{figure}
At the fixed point $x_1=y_1=x_2=y_2=0$, which we will write as $(0,0,0,0)$, or node $\emptyset$ in the invasion graph,
\begin{align*}
    J=\begin{pmatrix}
        1 & 0 & 0 & 0 \\
        0 & 0.75 & 0 & 0 \\
        0 & 0 & 1 & 0 \\
        0 & 0 & 0 & 0.5
    \end{pmatrix}.
\end{align*}
This is an unstable source and we record the signs of per-capita growth rates for all strains as $g(\emptyset)=(+,+,+,+)$. This node has an edge to every other node in the invasion graph.

Similarly node $x_1$ corresponds to the fixed point $x_1=10/11,y_1=x_2=y_2=0,$ or $(10/11,0,0,0)$. At this fixed point,
\begin{align*}
    J=\begin{pmatrix}
        -1 & -1/1.1 & 1/11 & 0 \\
        0 & 0.75-5/11 & 0 & 0 \\
        0 & 0 & 1+2/11 & 0 \\
        0 & 0 & 0 & 0.5
    \end{pmatrix}.
\end{align*}
It's an unstable saddle point and $g(x_1)=(0,+,+,+)$. This node has edges to all the other fixed points with $x_1$ present. Node $y_1$ corresponds to fixed point $(0,15/22,0,0)$ with $g(y_1)=(+,0,+,+)$. It is also an unstable saddle point like $x_1$ and has edges to all other fixed points with $y_1$ present.

\begin{table}[h]
\centering
\caption{Analysis results for the invasion graph}
\label{table}
\begin{tabular}{|c|c|c|c|} 
 \hline
  & & Jacobian & invasion growth \\ 
  Node & Fixed point & classification & rate signs \\
 \hline
 $\emptyset$ & $(0,0,0,0)$ & source & $(+,+,+,+)$ \\
 \hline
 $x_1$ & $(\frac{10}{11},0,0,0)$ & saddle & $(0,+,+,+)$ \\
 \hline
 $y_1$ & $(0,\frac{15}{22},0,0)$ & saddle & $(+,0,+,+)$ \\
 \hline
 $x_2$ & $(0,0,\frac{5}{6},0)$ & saddle & $(+,+,0,-)$ \\
 \hline
 $y_2$ & $(0,0,0,\frac{10}{9})$ & saddle & $(+,+,-,0)$ \\
 \hline
 $x_1y_1$ & $(\frac{35}{71},\frac{65}{142},0,0)$ & saddle & $(0,0,+,+)$ \\ 
 \hline
 $x_1x_2$ & $(1,0,1,0)$ & saddle & $(0,+,0,-)$ \\
 \hline
 $x_1y_2$ & $(\frac{10}{11},0,0,\frac{10}{9})$ & saddle & $(0,+,+,0)$ \\
 \hline
 $y_1x_2$ & $(0,\frac{15}{22},\frac{5}{6},0)$ & saddle & $(+,0,0,+)$ \\
 \hline
 $y_1y_2$ & $(0,\frac{31}{38},0,\frac{28}{19})$ & saddle & $(+,0,-,0)$ \\ 
 \hline
 $x_2y_2$ & $(0,0,\frac{5}{21},\frac{5}{7})$ & saddle & $(+,+,0,0)$ \\
 \hline
 $x_1y_1x_2$ & $(\frac{53}{83},\frac{65}{166},\frac{78}{83},0)$ & sink & $(0,0,0,-)$ \\
 \hline
 $x_1y_1y_2$ & $(\frac{5}{17},\frac{23}{34},0,\frac{24}{17})$ & sink & $(0,0,-,0)$ \\
 \hline
 $x_1x_2y_2$ & has negative values & saddle & $(0,+,0,0)$ \\
 \hline
 $y_1x_2y_2$ & has negative values & saddle & $(+,0,0,0)$ \\
 \hline
 $x_1y_1x_2y_2$ & $(\frac{1}{2},\frac{1}{2},\frac{1}{2},\frac{1}{2})$ & saddle & $(0,0,0,0)$ \\
 \hline
\end{tabular}
\end{table}

At node $x_2$, the fixed point is $(0,0,5/6,0)$. The Jacobian is 
\begin{align*}
    J=\begin{pmatrix}
        1+1/12 & 0 & 0 & 0 \\
        0 & 0.75 & 0 & 0 \\
        1/6 & 0 & -1 & -1/1.2 \\
        0 & 0 & 0 & 0.5-0.25/0.4
    \end{pmatrix}.
\end{align*} While this is also a saddle point, $g(x_2)=(+,+,0,-)$, node $x_2$ only has edges to $x_1x_2$, $y_1x_2$, and $x_1y_1x_2$ but not $x_2y_2$ or $x_1y_1x_2y_2$. Similarly $y_2$ corresponds to fixed point $(0,0,0,10/9)$ and $g(y_2)=(+,+,-,0)$, this node only goes to $x_1y_2$, $y_1y_2$, and $x_1y_1y_2$ but not $x_2y_2$ or $x_1y_1x_2y_2$.

We then move on to the next layer of the invasion graph to look at fixed points with 2 strains present. There are 6 such points. Above this layer, we have 2 fixed points with 3 strains present and the final layer has 1 fixed point with all strains present. We list all analysis results in table \ref{table}. The detailed verification of the edges in the invasion graph is listed in Appendix \ref{invasion-verf}. We can see from both the table and the invasion graph that there are two stable fixed points $(\frac{53}{83},\frac{65}{166},\frac{78}{83},0)$ and $(\frac{5}{17},\frac{23}{34},0,\frac{24}{17})$. 

With social interactions between the two hosts given by $C=\begin{pmatrix}
    0.1 & 0.1 \\
    0.2 & 0.2
\end{pmatrix}$, host 1 always has both strains present, and host 2 has one strain or the other. While this looks qualitatively the same as when the two hosts had no social interactions, the abundance profile is skewed from when $C$ is the zero matrix. When $C=0$, we would expect stable abundance profiles $(0.5,0.5,1,0)$ and $(0.5,0.5,0,2)$; with social interactions, strain $x$ becomes more abundant in host 1 than strain $y$ if host 2 only has strain $x$ present, and strain $y$ becomes more abundant in host 1 when host 2 only has strain $y$ present.

\subsection{Varying strength of social interaction}
To better examine the effect of the social interaction between the two hosts, we generalize the inter-host model and consider $C=m\begin{pmatrix}
    1 & 1 \\ 1 & 1
\end{pmatrix}$. Without loss of generality, we assume $0< m<2$. Our coupled system becomes
\begin{equation}
    \begin{cases}
        \dot x_1=x_1(1-x_1-y_1)+mx_1(x_2-x_1),\\
        \dot y_1=y_1(0.75-0.5x_1-y_1)+my_1(y_2-y_1),\\
        \dot x_2=x_2(1-x_2-y_2)+mx_2(x_1-x_2),\\
        \dot y_2=y_2(0.5-0.75x_2-0.25y_2)+my_2(y_1-y_2).
    \end{cases}
    \label{2-2-symm}
\end{equation}
The corresponding Jacobian and per-capita growth rates are
\begin{align*}
    &J=\\&\tiny\begin{pmatrix}
        1-2(1+m)x_1-y_1+mx_2 & -x_1 & mx_1 & 0 \\
        -0.5y_1 & 0.75-0.5x_1-2(1+m)y_1+my_2 & 0 & my_1\\
        mx_2 & 0 & 1-2(1+m)x_2-y_2+mx_1 & -x_2 \\
        0 & my_2 & -0.75y_2 & 0.5-0.75x_2-(0.5+2m)y_2+my_1
    \end{pmatrix}
\end{align*}
and 
$$r=\begin{pmatrix}
    1-(1+m)x_1-y_1+mx_2 \\ 0.75-0.5x_1-(1+m)y_1+my_2 \\ 1-(1+m) x_2-y_2+mx_1 \\ 0.5-0.75x_2-(0.25+m)y_2+my_1
\end{pmatrix}.$$
\begin{table}[h]
\centering
\caption{Fixed points under the general interaction matrix}
\label{tablem}
\begin{tabular}{|c|c|c|} 
 \hline
  Node & Fixed point & existence condition \\
 \hline
 $\emptyset$ & $(0,0,0,0)$ & None \\
 \hline
 $x_1$ & $(\frac{1}{m+1},0,0,0)$ & $m>0$ \\
 \hline
 $y_1$ & $(0,\frac{3}{4m+4},0,0)$ & $m>0$ \\
 \hline
 $x_2$ & $(0,0,\frac{1}{1+m},0)$ & $m>0$ \\
 \hline
 $y_2$ & $(0,0,0,\frac{2}{4m+1})$ & $m>0$ \\
 \hline
 $x_1y_1$ & $(\frac{4m+1}{4m^2+8m+2},\frac{3m+1}{4m^2+8m+2},0,0)$ & $m>0$ \\ 
 \hline
 $x_1x_2$ & $(1,0,1,0)$ & None \\
 \hline
 $x_1y_2$ & $(\frac{1}{m+1},0,0,\frac{2}{4m+1})$ & $m>0$ \\
 \hline
 $y_1x_2$ & $(0,\frac{3}{4m+4},\frac{1}{m+1},0)$ & $m>0$ \\
 \hline
 $y_1y_2$ & $(0,\frac{20m+3}{20m+4},0,\frac{5m+2}{5m+1})$ & $m>0$ \\ 
 \hline
 $x_2y_2$ & $(0,0,\frac{4m-1}{4m^2+5m-2},\frac{2m-1}{4m^2+5m-2})$ & $0<m<1/4$, or $m>1/2$ \\
 \hline
 $x_1y_1x_2$ & $(\frac{8m+1}{8m+2},\frac{2m+1}{8m^2+10m+2},\frac{8m^2+9m+2}{8m^2+10m+2},0)$ & $m>0$ \\
 \hline
 $x_1y_1y_2$ & $(\frac{1}{20m^2+16m+2},\frac{20m^2+15m+1}{20m^2+16m+2},0,\frac{10m^2+10m+2}{10m^2+8m+1})$ & $m>0$ \\
 \hline
 $x_1x_2y_2$ & $(\frac{8m^2+4m-2}{8m^2+3m-2},0,\frac{8m^2+4m-1}{8m^2+3m-2},\frac{-2m-1}{8m^2+3m-2})$ & $0<m<\frac{\sqrt{3}-1}{4}$ \\
 \hline
 $y_1x_2y_2$ & $(0,\frac{20m^2+11m-6}{20m^2+12m-8},\frac{-1}{5m^2+3m-2},\frac{5m-1}{5m-2})$ & $0<m<1/5$ \\
 \hline
 $x_1y_1x_2y_2$ & $(\frac{1}{2},\frac{1}{2},\frac{1}{2},\frac{1}{2})$ & None \\
 \hline
\end{tabular}
\end{table}

We list all fixed points and the corresponding analysis in tables \ref{tablem} and \ref{tablem1}. Here, node $x_2y_2$ only exists, i.e., the strain abundance values are positive, when $m\in(0,\frac{1}{4})\cup(\frac{1}{2},\infty)$. When it exists, it is a saddle point with growth rate signs $(+,+,0,0)$. The Jacobian analysis for sinks $x_1y_1x_2$ and $x_1y_1y_2$ can be numerically verified for $m\in(0,2)$. The fixed point $x_1x_2y_2$ exists only when $8m^2+4m-1<0$, or when $0<m<\frac{\sqrt{3}-1}{4}\approx0.183$. When it exists, it is a saddle point, which can be verified numerically. Its growth rate signs are $(0,+,0,0)$ as long as it exists. The saddle point $y_1x_2y_2$ exists when $0<m<\frac{1}{5}=0.2$. When it does exist, it has growth rate signs $(+,0,0,0)$ and the Jacobian analysis can be numerically verified. The Jacobian analysis for the fixed point $x_1y_1x_2y_2$ can also be numerically verified for $m\in(0,2)$.

\begin{table}[h!]
\centering
\caption{Stability analysis under varying social interaction strength}
\label{tablem1}
\begin{tabular}{|c|c|c|} 
 \hline
   & Jacobian & invasion growth \\ 
  Node  & classification & rate signs \\
 \hline
 $\emptyset$ & source & $(+,+,+,+)$ \\
 \hline
 $x_1$ & saddle & $(0,+,+,+)$ \\
 \hline
 $y_1$ & saddle & $(+,0,+,+)$ \\
 \hline
 $x_2$ & saddle & $(+,+,0,2m-1)$ \\
 \hline
 $y_2$ & saddle & $(+,+,4m-1,0)$ \\
 \hline
 $x_1y_1$ & saddle & $(0,0,+,+)$ \\ 
 \hline
 $x_1x_2$ & saddle & $(0,+,0,-)$ \\
 \hline
 $x_1y_2$ & saddle & $(0,+,8m^2+4m-1,0)$ \\
 \hline
 $y_1x_2$ & saddle & $(+,0,0,5m-1)$ \\
 \hline
 $y_1y_2$ & saddle & $(+,0,-,0)$ \\ 
 \hline
 $x_2y_2$ & saddle & $(+,+,0,0)$ \\
 \hline
 $x_1y_1x_2$ & sink & $(0,0,0,-)$ \\
 \hline
 $x_1y_1y_2$ & sink & $(0,0,-,0)$ \\
 \hline
 $x_1x_2y_2$ & saddle & $(0,+,0,0)$ \\
 \hline
 $y_1x_2y_2$ & saddle & $(+,0,0,0)$ \\
 \hline
 $x_1y_1x_2y_2$ & saddle & $(0,0,0,0)$ \\
 \hline
\end{tabular}
\end{table}

The cosine similarity of the abundance profiles of the two hosts without social interactions would be 
\begin{align*}
S_c(\langle0.5,0.5\rangle,\langle1,0\rangle)=\frac{\langle0.5,0.5\rangle\cdot\langle1,0\rangle}{0.5\sqrt{2}}=\frac{1}{\sqrt{2}}, \\
S_c(\langle0.5,0.5\rangle,\langle0,2\rangle)=\frac{\langle0.5,0.5\rangle\cdot\langle0,2\rangle}{0.5\sqrt{2}\times 2}=\frac{1}{\sqrt{2}}.
\end{align*}
And with social interaction strength depending on $m$, the cosine similarity becomes
\begin{align*}
    S_c(\langle\frac{8m+1}{8m+2},\frac{2m+1}{8m^2+10m+2}\rangle,\langle\frac{8m^2+9m+2}{8m^2+10m+2},0\rangle)\\
    =S_c(\langle(8m+1)(m+1),2m+1\rangle,\langle1,0\rangle)\\
    =\frac{8m^2+9m+1}{\sqrt{(8m^2+9m+1)^2+(2m+1)^2}}>\frac{1}{\sqrt{2}},
\end{align*}
\begin{align*}
    S_c(\langle\frac{1}{20m^2+16m+2},\frac{20m^2+15m+1}{20m^2+16m+2}\rangle,\langle0,\frac{10m^2+10m+2}{10m^2+8m+1}\rangle)\\
    =S_c(\langle1,20m^2+15m+1\rangle,\langle0,1\rangle)\\
    =\frac{20m^2+15m+1}{\sqrt{(20m^2+15m+1)^2+(1)^2}}>\frac{1}{\sqrt{2}}.
\end{align*}
Both cosine similarities are now greater and increase as $m$ increases.

\subsection{Variation on in-host dynamics}
\begin{figure}[h]
    \centering
    \begin{subfigure}{0.45\textwidth}
        \includegraphics[width=\textwidth]{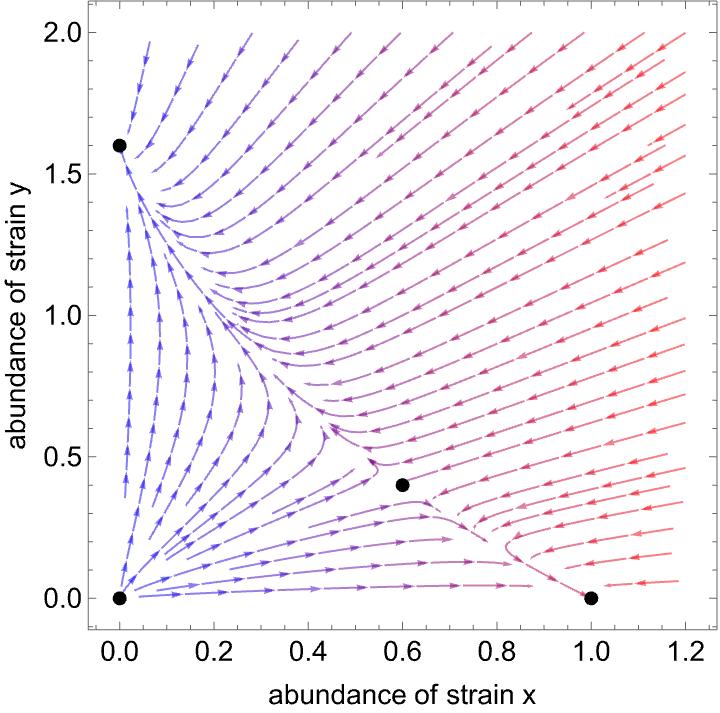}
        \caption{phase portrait}
    \end{subfigure}
    \hfill
    \begin{subfigure}{0.45\textwidth}
        \centering
        \begin{tikzpicture}[-latex, auto, node distance = 0.3 cm and 0.3 cm, on grid, semithick, state/.style ={circle, top color =white, bottom color = processblue!20, draw, processblue, text=blue, minimum width =0.3 cm}]
        \node[state] (A) at (2, 5.3) {$xy$};
        \node[state] (B) at (0, 3.3) {$x$};
        \node[state] (C) at (2, 1.3) {$\emptyset$};
        \node[state] (D) at (4, 3.3) {$y$};
        \node[] (E) at (0,0) {};
        \path (A) edge (B) edge (D);
        \path (C) edge (A) edge (B) edge (D);
        \end{tikzpicture}
        \caption{invasion graph}
    \end{subfigure}
    \caption{Modified microbial dynamics in host 1}
    \label{1-new}
\end{figure}
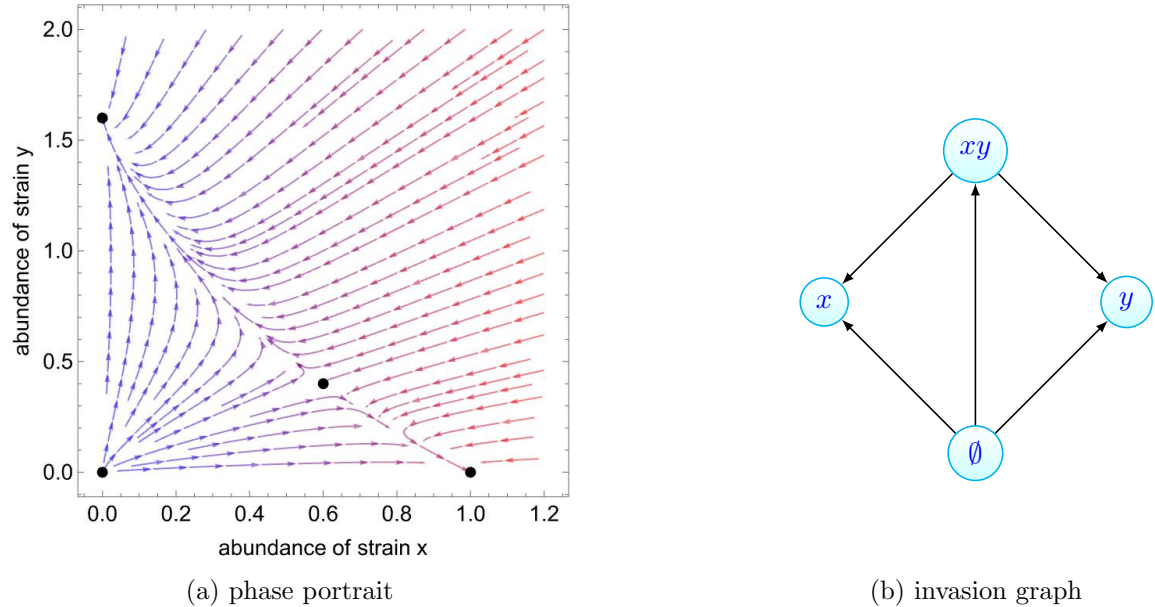

Host 1 had a stable coexistence between strain $x$ and strain $y$ in the two previous examples and the social interaction with host 2 didn't qualitatively change that. We would now like to see what happens if both host 1 and host 2 have unstable coexistence. We keep the governing equations for host 2 the same, and modify the equations for host 1 to
\begin{equation}
    \dot x=x(1-x-y),\quad \dot y=y(0.8-x-0.5y).
    \label{host1unstable}
\end{equation}
Now the phase portrait of the microbial dynamics in host 1 and the corresponding invasion graph are shown in figure \ref{1-new}. Host 1 now has an unstable coexistence at $(0.6,0.4)$, an unstable source at $(0,0)$, and two stable sinks $(1,0)$ and $(0,1.6)$. The invasion graph looks the same as host 2 without social interactions.

The modified coupled system with symmetric and varying strength social interaction is
\begin{equation}
    \begin{cases}
        \dot x_1=x_1(1-x_1-y_1)+mx_1(x_2-x_1),\\
        \dot y_1=y_1(0.8-x_1-0.5y_1)+my_1(y_2-y_1),\\
        \dot x_2=x_2(1-x_2-y_2)+mx_2(x_1-x_2),\\
        \dot y_2=y_2(0.5-0.75x_2-0.25y_2)+my_2(y_1-y_2).
    \end{cases}
    \label{2-2-uu}
\end{equation}
Its corresponding Jacobian and per-capita growth rates are
\begin{align*}
    &J=\\&\tiny\begin{pmatrix}
        1-2(1+m)x_1-y_1+mx_2 & -x_1 & mx_1 & 0 \\
        -y_1 & 0.8-x_1-(1+2m)y_1+my_2 & 0 & my_1\\
        mx_2 & 0 & 1-2(1+m)x_2-y_2+mx_1 & -x_2 \\
        0 & my_2 & -0.75y_2 & 0.5-0.75x_2-(0.5+2m)y_2+my_1
    \end{pmatrix}
\end{align*}
and 
$$r=\begin{pmatrix}
    1-(1+m)x_1-y_1+mx_2 \\ 0.8-x_1-(0.5+m)y_1+my_2 \\ 1-(1+m) x_2-y_2+mx_1 \\ 0.5-0.75x_2-(0.25+m)y_2+my_1
\end{pmatrix}.$$

\begin{table}
\caption{Fixed points for unstable coexistence in both hosts}
\label{tableu}
\begin{tabular}{|c|c|c|} 
 \hline
  Node & Fixed point & existence condition \\
 \hline
 $\emptyset$ & $(0,0,0,0)$ & None \\
 \hline
 $x_1$ & $(\frac{1}{m+1},0,0,0)$ & $m>0$ \\
 \hline
 $y_1$ & $(0,\frac{8}{10m+5},0,0)$ & $m>0$ \\
 \hline
 $x_2$ & $(0,0,\frac{1}{1+m},0)$ & $m>0$ \\
 \hline
 $y_2$ & $(0,0,0,\frac{2}{4m+1})$ & $m>0$ \\
 \hline
 $x_1y_1$ & $(\frac{10m-3}{10m^2+15m-5},\frac{8m-2}{10m^2+15m-5},0,0)$ & $0<m<\frac{1}{4}$, or $m>\frac{3}{10}$ \\ 
 \hline
 $x_1x_2$ & $(1,0,1,0)$ & None \\
 \hline
 $x_1y_2$ & $(\frac{1}{m+1},0,0,\frac{2}{4m+1})$ & $m>0$ \\
 \hline
 $y_1x_2$ & $(0,\frac{8}{10m+5},\frac{1}{m+1},0)$ & $m>0$ \\
 \hline
 $y_1y_2$ & $(0,\frac{52m+8}{30m+5},0,\frac{52m+10}{30m+5})$ & $m>0$ \\ 
 \hline
 $x_2y_2$ & $(0,0,\frac{4m-1}{4m^2+5m-2},\frac{2m-1}{4m^2+5m-2})$ & $0<m<\frac{1}{4}$, or $m>\frac{1}{2}$ \\
 \hline
 $x_1y_1x_2$ & $(\frac{20m^2+12m-3}{20m^2+10m-5},\frac{-4m-2}{20m^2+10m-5},\frac{20m^2+12m-5}{20m^2+10m-5},0)$ & $0<m<\frac{2\sqrt{6}-3}{10}\approx0.19$ \\
 \hline
 $x_1y_1y_2$ & $(\frac{-22m-3}{30m^2-5m-5},\frac{52m^2+20m-2}{30m^2-5m-5},0,\frac{52m^2+22m-10}{30m^2-5m-5})$ &  $0<m<\frac{\sqrt{51}-5}{26}\approx0.082$\\
 \hline
 $x_1x_2y_2$ & $(\frac{8m^2+4m-2}{8m^2+3m-2},0,\frac{8m^2+4m-1}{8m^2+3m-2},\frac{-2m-1}{8m^2+3m-2})$ & $0<m<\frac{\sqrt{3}-1}{4}\approx0.183$ \\
 \hline
 $y_1x_2y_2$ & $(0,\frac{52m^2+30m-16}{30m^2+5m-10},\frac{-22m-5}{30m^2+5m-10},\frac{52m^2+32m-5}{30m^2+5m-10})$ & $0<m<\frac{\sqrt{129}-8}{26}\approx0.129$ \\
 \hline
 $x_1y_1x_2y_2$ & $(\frac{44m^2+30m-6}{80m^2+55m-10},\frac{36m^2+26m-4}{80m^2+55m-10},\frac{44m^2+30m-5}{80m^2+55m-10},\frac{36m^2+24m-5}{80m^2+55m-10})$ & $0<m<\frac{\sqrt{313}-13}{36}\approx0.130$ \\
 & & or $m>\frac{1}{6}\approx0.167$ \\
 \hline
\end{tabular}
\end{table}

We find all fixed points and their existence conditions and list them in table \ref{tableu}. We verify the eigenvalues of the Jacobian and invasion growth rate signs and list the results in table \ref{tableu1}. Results for fixed points $x_1y_1$, $y_1y_2$, $x_2y_2$, $x_1y_1x_2$, $x_1y_1y_2$, $x_1x_2y_2$, $y_1x_2y_2$, $x_1y_1x_2y_2$ are numerically verified on $m\in(0,2)$ where the fixed point exists.

\begin{table}
\centering
\caption{Stability analysis for unstable coexistence in both hosts}
\label{tableu1}
\begin{tabular}{|c|c|c|} 
 \hline
   & Jacobian & invasion growth \\ 
  Node & classification & rate signs \\
 \hline
 $\emptyset$ & source & $(+,+,+,+)$ \\
 \hline
 $x_1$ & saddle & $(0,4m-1,+,+)$ \\
 \hline
 $y_1$ & saddle & $(10m-3,0,+,+)$ \\
 \hline
 $x_2$ & saddle & $(+,+,0,2m-1)$ \\
 \hline
 $y_2$ & saddle & $(+,+,4m-1,0)$ \\
 \hline
 $x_1y_1$ & saddle & $(0,0,+,+)$ \\ 
 \hline
 $x_1x_2$ & sink & $(0,-,0,-)$ \\
 \hline
 $x_1y_2$ & $0<m<\frac{\sqrt{51}-5}{26}$ sink & $(0,-,-,0)$ \\
  & $\frac{\sqrt{51}-5}{26}<m<\frac{\sqrt{3}-1}{4}$ saddle & $(0,+,-,0)$ \\
  & $m>\frac{\sqrt{3}-1}{4}$ saddle & $(0,+,+,0)$ \\
 \hline
 $y_1x_2$ & $0<m<\frac{\sqrt{129}-8}{26}$ sink & $(-,0,0,-)$ \\
  & $\frac{\sqrt{129}-8}{26}<m<\frac{2\sqrt{6}-3}{10}$ saddle & $(-,0,0,+)$ \\
  & $m>\frac{2\sqrt{6}-3}{10}$ saddle & $(+,0,0,+)$ \\
 \hline
 $y_1y_2$ & sink & $(-,0,-,0)$ \\ 
 \hline
 $x_2y_2$ & saddle & $(+,+,0,0)$ \\
 \hline
 $x_1y_1x_2$ & $0<m<\frac{1}{6}$ saddle & $(0,0,0,-)$ \\
    & $\frac{1}{6}<m<\frac{2\sqrt{6}-3}{10}$ saddle & $(0,0,0,+)$ \\
 \hline
 $x_1y_1y_2$ & saddle & $(0,0,-,0)$ \\
 \hline
 $x_1x_2y_2$ & $0<m<\frac{\sqrt{313}-13}{36}$ saddle & $(0,0,-,0)$ \\
  & $\frac{\sqrt{313}-13}{36}<m<\frac{\sqrt{3}-1}{4}$ saddle & $(0,0,+,0)$ \\
 \hline
 $y_1x_2y_2$ & saddle & $(-,0,0,0)$ \\
 \hline
 $x_1y_1x_2y_2$ & saddle & $(0,0,0,0)$ \\
 \hline
\end{tabular}
\end{table}

When host 1 had stable coexistence, we had two sinks, $x_1y_1x_2$ and $x_1y_1x_2$. Now that host 1 has unstable coexistence, we have four sinks, $x_1x_2,x_1y_2,y_1x_2,y_1y_2$, which are the same sinks we would have if there was no social interaction between the two hosts. We need to point out that $x_1y_2$ and $y_1x_2$ are only sinks when $m$ is small, less than 0.082 and 0.129, respectively. When $m$ is greater than 0.129, $x_1y_2$ and $y_1x_2$ become saddles, and the only sinks left are $x_1x_2$ and $y_1y_2$. In this situation, the cosine similarity between the abundance profile of the two hosts is 1. Without social interactions, the cosine similarity of the abundance profiles of the two hosts could be 1 or 0 depending on whether they have the same or different strains of microbes present.

Now that both hosts have unstable coexistence, when $m$ reaches a critical threshold, 0.129, two of the four sinks become saddles and the dynamics changes. The attraction basins for the sinks change, and the two hosts "sync" their microbial profile. We would like to point out another difference between the two scenarios where host 1 has stable coexistence and host 1 has unstable coexistence. In the first scenario, with social interactions, host 1 still has coexistence and host 2 has either strain $x$ or strain $y$. While the abundance profile in host 1 is skewed under social interaction with host 2, once they stop their social interaction, each host will go back to their corresponding steady states, i.e. $(0.5,0.5)$ for host 1 and either $(1,0)$ or $(0,2)$ for host 2. In other words, there is no lasting effect on the individual microbial profiles from the social interaction. On the other hand, when both hosts have unstable coexistence, with strong social interactions, $m>0.129$, the microbial profiles will "sync" so both hosts will have the same strain of microbe present. And after the social interactions disappear they will stay "in sync" as they will have the same strain of microbe present. Even if they had different strains present before the social interaction, the social interaction pushes the dynamics out of one attraction basin into another, so that once they reach the new steady state, even if we remove the social interaction, the two hosts stay "in sync" in their microbial profile. In this case, the social interaction has a lasting effect on individual microbial profiles.

\begin{figure}
    \centering
    \begin{subfigure}[b]{0.45\textwidth}
        \includegraphics[width=\textwidth]{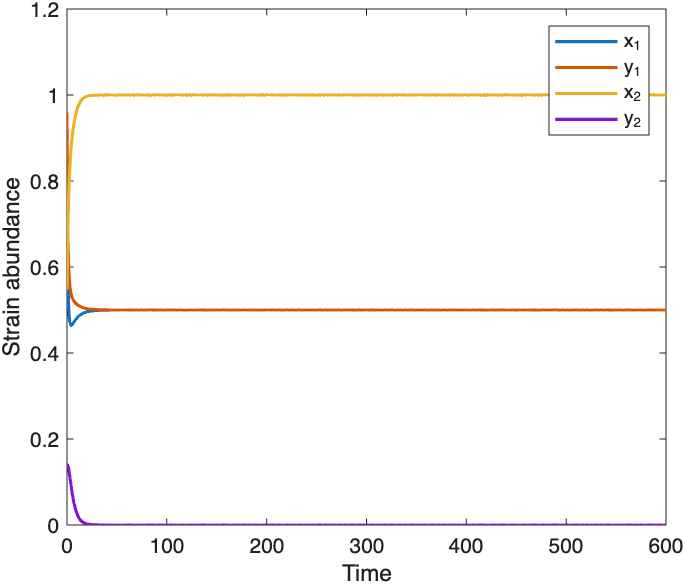}
        \caption{$m=0$}
    \end{subfigure}
    \hfill
    \begin{subfigure}[b]{0.45\textwidth}
        \includegraphics[width=\textwidth]{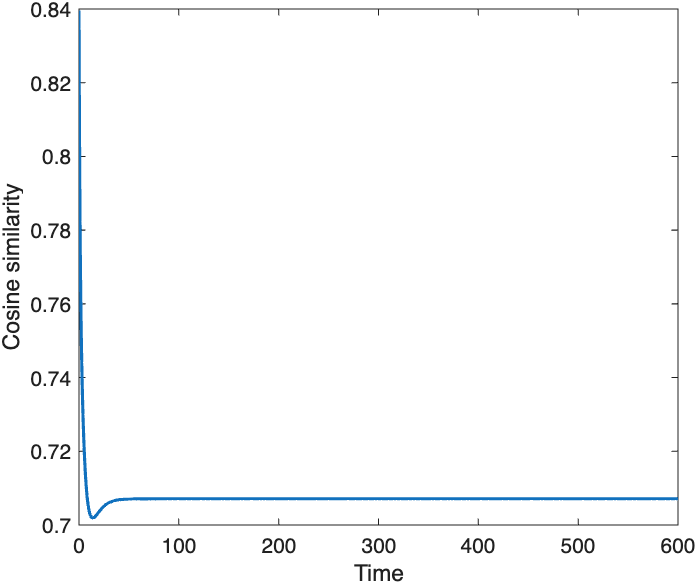}
        \caption{$m=0$}
    \end{subfigure}\\
    \begin{subfigure}[b]{0.45\textwidth}
        \includegraphics[width=\textwidth]{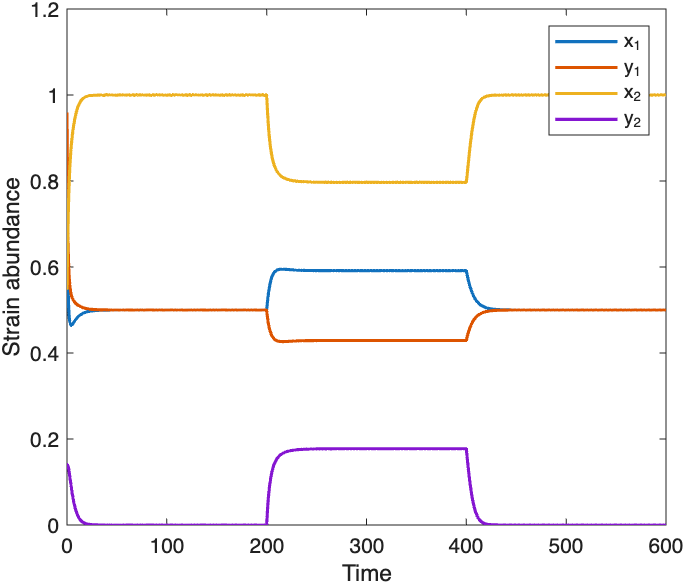}
        \caption{$m=0.1$}
    \end{subfigure}
    \hfill
    \begin{subfigure}[b]{0.45\textwidth}
        \includegraphics[width=\textwidth]{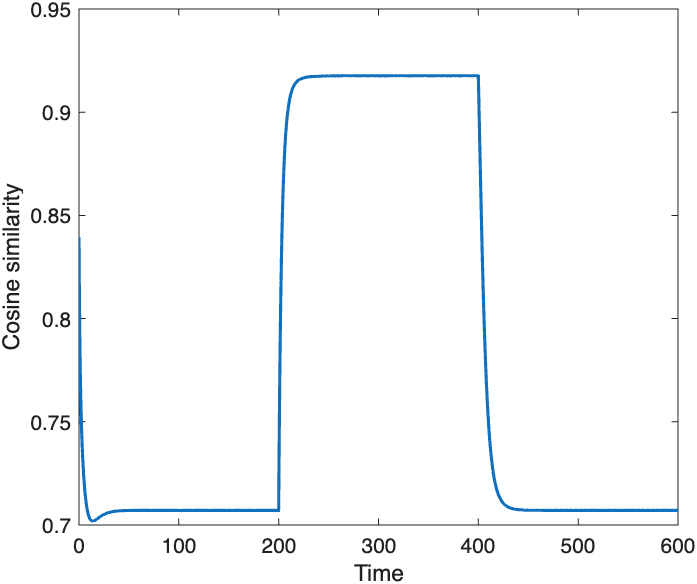}
        \caption{$m=0.1$}
    \end{subfigure}\\
    \begin{subfigure}[b]{0.45\textwidth}
        \includegraphics[width=\textwidth]{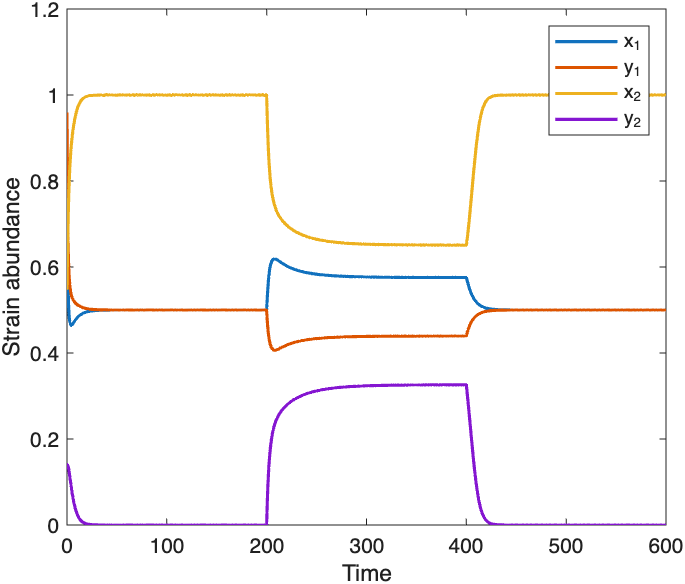}
        \caption{$m=0.2$}
    \end{subfigure}
    \hfill
    \begin{subfigure}[b]{0.45\textwidth}
        \includegraphics[width=\textwidth]{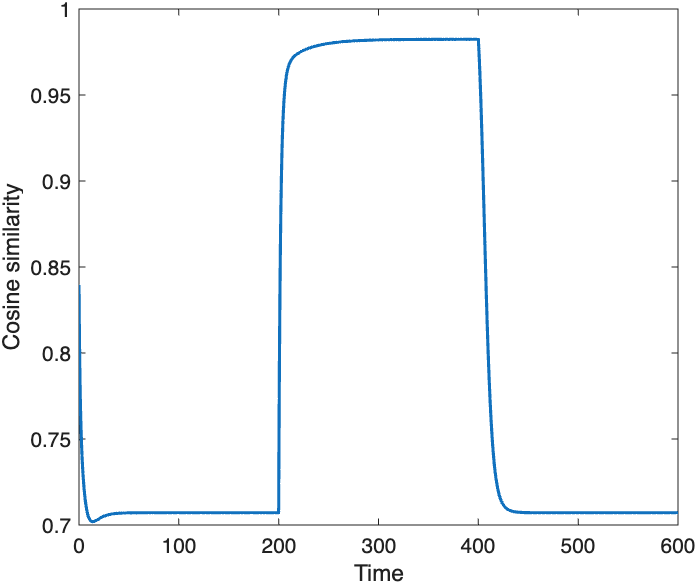}
        \caption{$m=0.2$}
    \end{subfigure}
    \caption{Microbial dynamics in host 1 and host 2 when host 1 has stable coexistence}
    \label{coexistence1}
\end{figure}

To better illustrate the lasting effect of social interaction on the microbial dynamics, we plot the time evolution of the microbial dynamics in the two different scenarios under different $m$ values. In Figure \ref{coexistence1}, (a) (c) (e) are the abundance profiles of the two hosts and (b) (d) (f) show the cosine similarity between the two hosts' abundance profiles. From time 0 to 200 the two hosts have no social interaction, from time 200 to 400 they have social interaction with strength $m$, and from time 400 to 600, the social interaction is removed. We can see without social interaction, $x_1$ and $y_1$ always converge to 0.5 and with the randomly generated initial condition $(0.8909,0.9593,0.5472,0.1386)$ here, $y_2\to0$ and $x_2\to1$. With different initial conditions, we could have $x_2\to0,y_2\to2$ as well. The cosine similarity of the abundance profiles always stabilizes around $1/\sqrt{2}\approx0.7$, and it increases with stronger social interaction. We can see when host 1 has stable coexistence between the two strains of microbes, there is no lasting effect from social interaction between the two hosts. They each return to their original steady state once the social interaction stops. We would like to point out that from time 200 to 400, in (c) and (e) it looks like the system has reached an equilibrium that's not in table \ref{tablem}, this could be because we reached some slow converging submanifold in the phase space.

\begin{figure}
    \centering
    \begin{subfigure}[b]{0.45\textwidth}
        \includegraphics[width=\textwidth]{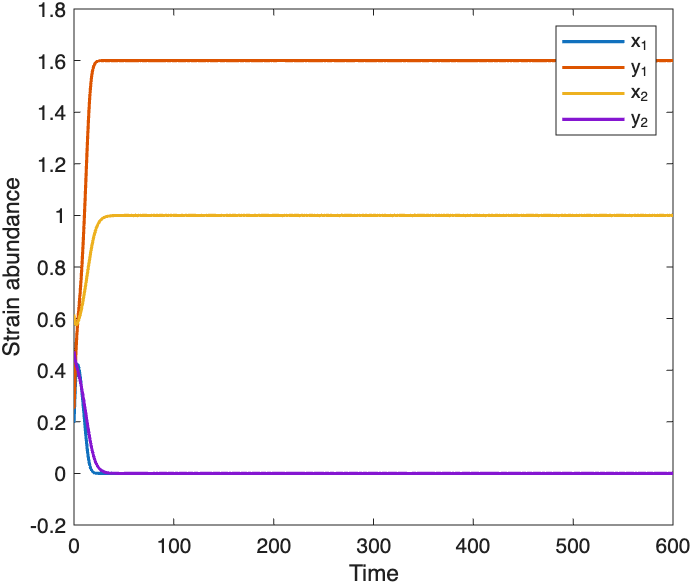}
        \caption{$m=0$}
    \end{subfigure}
    \hfill
    \begin{subfigure}[b]{0.45\textwidth}
        \includegraphics[width=\textwidth]{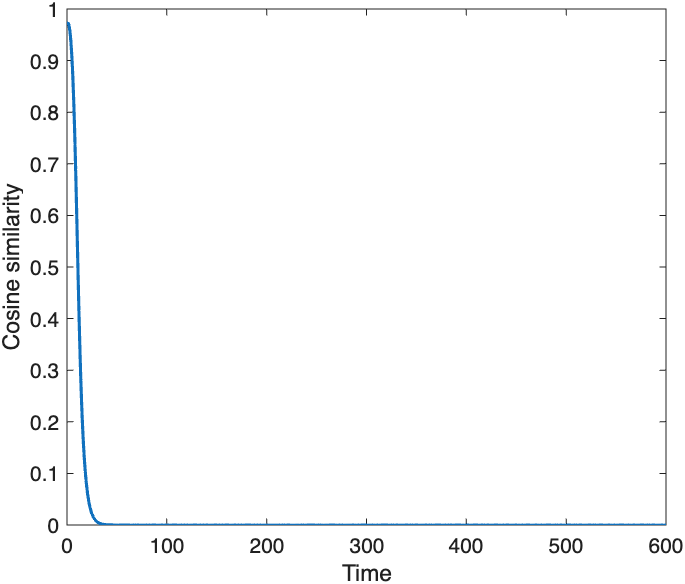}
        \caption{$m=0$}
    \end{subfigure}\\
    \begin{subfigure}[b]{0.45\textwidth}
        \includegraphics[width=\textwidth]{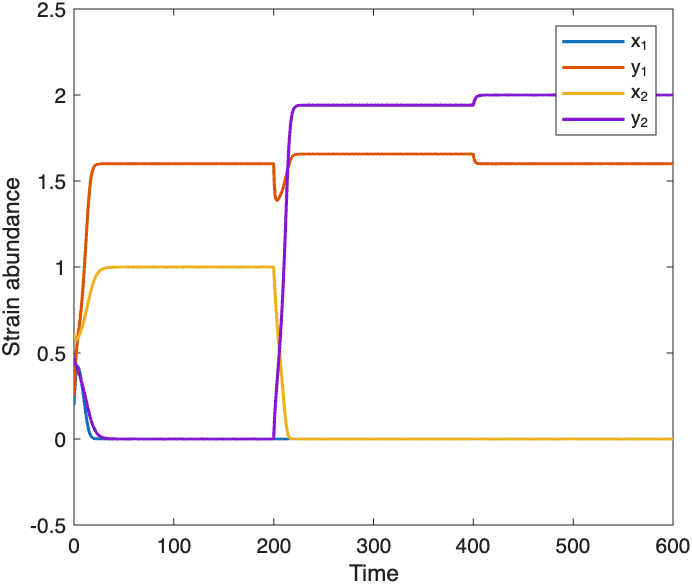}
        \caption{$m=0.1$}
    \end{subfigure}
    \hfill
    \begin{subfigure}[b]{0.45\textwidth}
        \includegraphics[width=\textwidth]{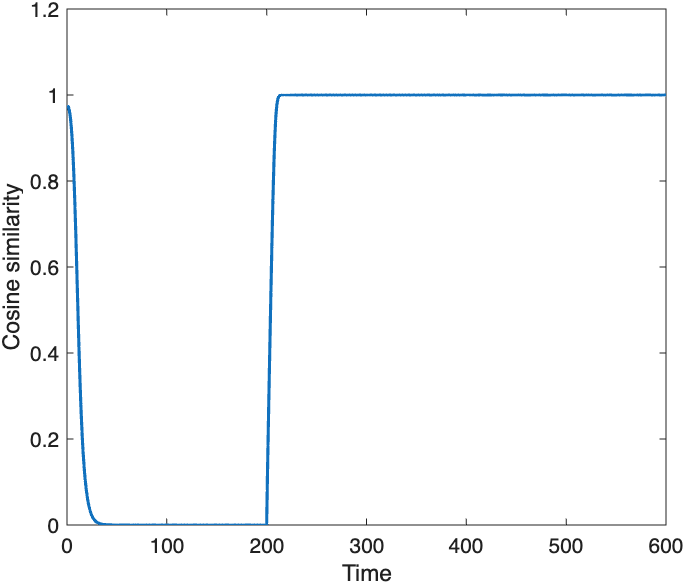}
        \caption{$m=0.1$}
    \end{subfigure}\\
    \begin{subfigure}[b]{0.45\textwidth}
        \includegraphics[width=\textwidth]{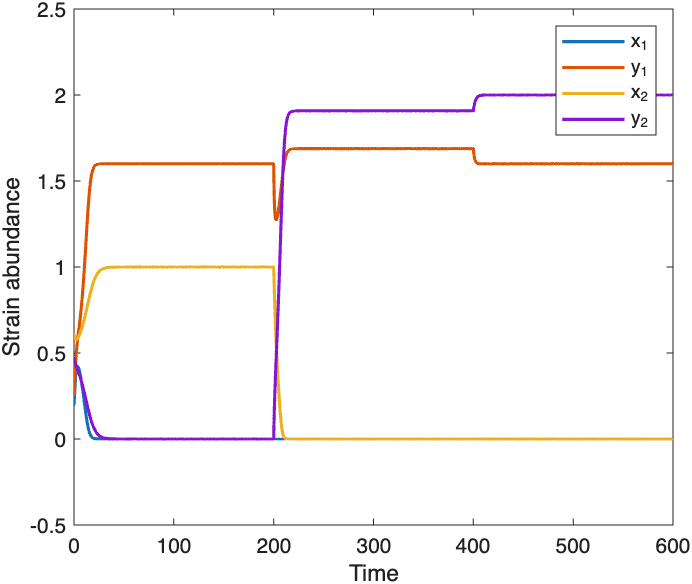}
        \caption{$m=0.2$}
    \end{subfigure}
    \hfill
    \begin{subfigure}[b]{0.45\textwidth}
        \includegraphics[width=\textwidth]{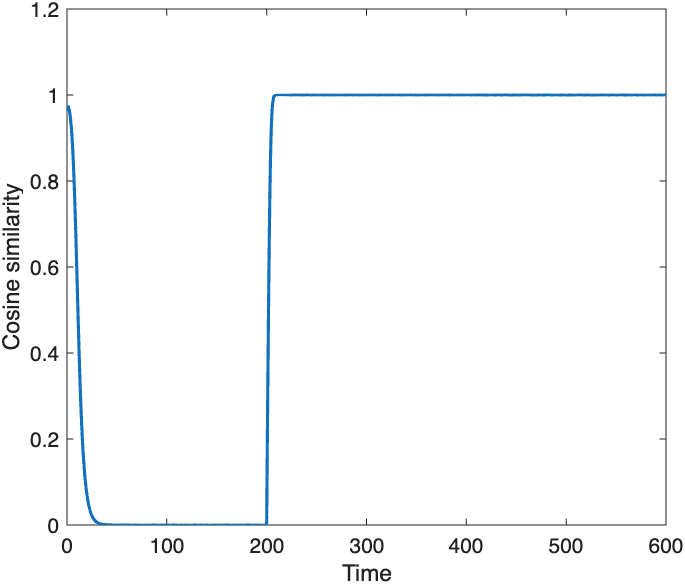}
        \caption{$m=0.2$}
    \end{subfigure}
    \caption{Microbial dynamics in host 1 and host 2 when host 1 has unstable coexistence}
    \label{unstable1}
\end{figure}

For comparison, we plot the same evolution in both hosts when host 1 has unstable coexistence between the two strains of microbes, see figure \ref{unstable1}. Here with a randomly generated initial condition $(0.1966,0.2511,0.6160,0.4733)$ and no social interactions $x_1\to0,y_1\to1.6,x_2\to1,y_2\to0$. With no social interaction, the cosine similarity decreases to 0. However, with social interaction, even when $m=0.1$, we can see that the system is pushed out of the original attraction basin of the fixed point $y_1x_2$. Now, the system converges to $y_1y_2$. This also forces the cosine similarity to converge to 1 instead of 0. When $m=0.2$ this process happens even faster. They reach the same steady states. Referring to tables \ref{tableu} and \ref{tableu1}, we can see the fixed point $y_1x_2$ exists as long as $m>0$ and is a local sink when $m<0.129$. Here we already escaped the sink when $m=0.1<0.129$. This could be because when $m=0.1$ the attraction basin of $y_1x_2$ is already very small and the steady state $(0,1.6,1,0)$ is outside of it.

It is possible to start with an initial condition where host 1 and host 2 would have the same strain of microbe present without social interactions. In this case adding or removing social interaction doesn't really affect their microbial profiles much.

This simple example with only 2 strains of microbes in 2 hosts already displays rich dynamics where social interaction between the hosts could have lasting effects on their microbial profiles. The social interaction can push the system out of the attraction basin of one local sink and into another. The number of sinks and their attraction basin depend on both the parameters of the original individual microbial dynamics and the social interaction strength $m$. 

Judging by the equations of individual dynamics for host 1 and 2, we would like to speculate that when the in-host competition is weak we have stable coexistence between different strains, a global sink, and social interaction doesn't have a lasting effect on the individual microbial dynamics. When the in-host competition is strong we have unstable coexistence and typically multiple local sinks, in which case social interaction can have lasting effects on the individual microbial dynamics.

\section{Numerical Simulation}
\label{num}
In the earlier examples with 2 strains of microbes in 2 hosts, when we couple the dynamics to model host social interactions, we had to solve a 4-variable system to find 16 fixed points. With varying strength of social interaction, the fixed point and Jacobian analysis got quite involved. In  general, with more strains of microbes in each host, and more number of hosts, finding all fixed points and analyzing the Jacobian systematically seem impractical. We will explore these scenarios with numerical simulations mostly.

\subsection{2 hosts, each with 10 strains of microbes}
In the first numerical experiment, we consider 2 hosts, each with 10 strains of microbes inside. The dynamics is governed by
\begin{equation}
    \frac{dN_i}{dt}=r_iN_i\left(1-\frac{\sum_{j=1}^{10}\beta_{ij}N_j}{K_i}\right), \quad i=1,\dots,10.
    \label{2-10-ind}
\end{equation}
Here $r_i$ is the growth rate of each strain, which will be drawn from a uniform distribution on $(0.1,0.4)$. $K_i$ is the carrying capacity and will be drawn from a uniform distribution $U(0.8,1.2)$. The competition matrix $B=(\beta_{ij})$ is a $10\times10$ matrix, with $\beta_{ii}=1$, and $\beta_{ij}$ drawn from $U(0.8,1.6)$ when $i\neq j$ to ensure strong competition. We choose $\beta_{ij}$ so that coexistence is probably unstable and social interaction is more likely to have a lasting effect on individual microbial profiles.

Without social interactions, we have 2 independent systems of 10 equations. At fixed points, for each $i$, we have either
\begin{equation}
    N_i=0
    \label{extinct}
\end{equation}
or
\begin{equation}
    \sum_{i=1}^{10}\beta_{ij}N_j=K_i
    \label{coexist}
\end{equation}

Assuming $B$ and all principle submatrices of $B$ are invertible, we have 1 fixed point where all strains are extinct, i.e., $N_i=0$ for all $i$; 10 fixed points with one strain absent, i.e., $N_i=0$ for one $i$ and \eqref{coexist} for the rest; in total, we have $C_{10}^0+C_{10}^1+C_{10}^2+\dots+C_{10}^{10}=2^{10}$ fixed points.

The Jacobian of an individual system is
{\scriptsize
\begin{align*}
    &J=\\
    &\tiny\begin{pmatrix}
        r_1\left(1-\dfrac{\sum_{j=1}^{10}\beta_{1j}N_j}{K_1}\right)-r_1N_1\dfrac{\beta_{11}}{K_1} & -r_1N_1\dfrac{\beta_{12}}{K_1} & \dots & -r_1N_1\dfrac{\beta_{110}}{K_1}\\
        -r_2N_2\dfrac{\beta_{21}}{K_2} & r_2\left(1-\dfrac{\sum_{j=1}^{10}\beta_{2j}N_j}{K_2}\right)-r_2N_2\dfrac{\beta_{22}}{K_2} & \dots & -r_2N_2\dfrac{\beta_{210}}{K_2}\\
        \vdots & \vdots & \ddots & \vdots\\
        -r_{10}N_{10}\dfrac{\beta_{101}}{K_{10}} & -r_{10}N_{10}\dfrac{\beta_{102}}{K_{10}} & \dots & r_{10}\left(1-\dfrac{\sum_{j=1}^{10}\beta_{10j}N_j}{K_{10}}\right)-r_{10}N_{10}\dfrac{\beta_{1010}}{K_{10}}
    \end{pmatrix}.
\end{align*}}

At the fixed point where all strains are extinct, i.e., $N_i=0$ for all $i$, the Jacobian becomes
\begin{align*}
    J=\begin{pmatrix}
        r_1 & 0 & \dots & 0 \\
        0 & r_2 & \dots & 0 \\
        \vdots & \vdots & \ddots & \vdots \\
        0 & 0 & \dots & r_{10}
    \end{pmatrix}.
\end{align*}
Obviously, this is an unstable source where each and every eigenvalue is positive.

At the fixed point where all strains are present, i.e., $N_i>0,i=1,\dots,10$, 
\begin{align*}
    J=\begin{pmatrix}
        -r_1N_1\dfrac{\beta_{11}}{K_1} & -r_1N_1\dfrac{\beta_{12}}{K_1} & \dots & -r_1N_1\dfrac{\beta_{110}}{K_1} \\
        -r_2N_2\dfrac{\beta_{21}}{K_2} & -r_2N_2\dfrac{\beta_{22}}{K_2} & \dots & -r_2N_2\dfrac{\beta_{210}}{K_2} \\
        \vdots & \vdots & \ddots & \vdots \\
        -r_{10}N_{10}\dfrac{\beta_{101}}{K_{10}} & -r_{10}N_{10}\dfrac{\beta_{102}}{K_{10}} & \dots & -r_{10}N_{10}\dfrac{\beta_{1010}}{K_{10}}
    \end{pmatrix}
\end{align*}
is a negative matrix. It has at least one negative eigenvalue. So the fixed point with every strain present could be either asymptotically stable, or an unstable saddle point.

To study the stability of the 10 fixed points with one strain absent, i.e., $N_i=0$ for some $i$, we can just look at the fixed point with $N_{10}=0$, since we can always permute or relabel the strains. In this case, we have
\begin{align*}
    J=\begin{pmatrix}
        -r_1N_1\dfrac{\beta_{11}}{K_1} & -r_1N_1\dfrac{\beta_{12}}{K_1} & \dots & -r_1N_1\dfrac{\beta_{110}}{K_1} \\
        -r_2N_2\dfrac{\beta_{21}}{K_2} & -r_2N_2\dfrac{\beta_{22}}{K_2} & \dots & -r_2N_2\dfrac{\beta_{210}}{K_2} \\
        \vdots & \vdots & \ddots & \vdots \\
        0 & 0 & \dots & r_{10}\left(1-\dfrac{\sum_{j=1}^9\beta_{10j}N_j}{K_{10}}\right)
    \end{pmatrix}.
\end{align*}

When we have $p$ strains present and $10-p$ strains absent, under permutation, we can always assume strains $1,2,\dots, p$ are present and strains $p+1,p+2,\dots,10$ are absent. The Jacobian in these cases look like
\begin{align*}
    J&=\begin{pmatrix}
        A & C \\ 0 & D
    \end{pmatrix},\\
    A&=\begin{pmatrix}
        -r_1N_1\dfrac{\beta_{11}}{K_1} & \dots & -r_1N_1\dfrac{\beta_{1p}}{K_1} \\
        \vdots & \ddots & \vdots \\
        -r_pN_p\dfrac{\beta_{p1}}{K_p} & \dots & -r_pN_p\dfrac{\beta_{pp}}{K_p}
    \end{pmatrix},
    C=\begin{pmatrix}
        -r_1N_1\dfrac{\beta_{1p+1}}{K_1} & \dots & -r_1N_1\dfrac{\beta_{110}}{K_1} \\
        \vdots & \ddots & \vdots \\
        -r_pN_p\dfrac{\beta_{pp+1}}{K_p} & \dots & -r_pN_p\dfrac{\beta_{p10}}{K_p}
    \end{pmatrix}, \\D&=\begin{pmatrix}
        r_{p+1}\left(1-\dfrac{\sum_{j=1}^p\beta_{p+1j}N_j}{K_{p+1}}\right) & \dots & 0 \\
        \vdots & \ddots & \vdots \\
        0 & \dots & r_{10}\left(1-\dfrac{\sum_{j=1}^p\beta_{10j}N_j}{K_{10}}\right)
    \end{pmatrix}.
\end{align*}
The eigenvalues of $J$ are $\sigma(J)=\sigma(A)\cup\sigma(D)$, the combination of eigenvalues of $A$ and $D$. $D$ is diagonal, so the eigenvalues are its diagonal entries. $A$ is a negative matrix, so it has at least one negative eigenvalue.

Now consider regular social interaction between the two hosts. The 2 individual systems will be coupled in the following way.
\begin{equation}
    \begin{cases}
        \dfrac{dN_i^{(1)}}{dt}&=r_i^{(1)}N_i^{(1)}\left(1-\dfrac{\sum_{j=1}^{10}\beta_{ij}^{(1)}N_j^{(1)}}{K_i^{(1)}}\right)+m_i^{(1)}\left(N_i^{(2)}-N_i^{(1)}\right),\\
        \dfrac{dN_i^{(2)}}{dt}&=r_i^{(2)}N_i^{(2)}\left(1-\dfrac{\sum_{j=1}^{10}\beta_{ij}^{(2)}N_j^{(2)}}{K_i^{(2)}}\right)+m_i^{(2)}\left(N_i^{(1)}-N_i^{(2)}\right),\\
        &i=1,\dots,10.
    \end{cases}
    \label{2-10-soc}
\end{equation}

For simplicity, we let all $m_i^{(j)}=m$ be the same. Let $J^{(1)}$ be the Jacobian for host 1 and $J^{(2)}$ be the Jacobian for host 2, the Jacobian of the coupled system is
\begin{align*}
    J=\begin{pmatrix}
        J^{(1)}-mI & mI \\
        mI & J^{(2)}-mI
    \end{pmatrix}.
\end{align*}

To study the effect of $m$, strength of the social interaction, on long term individual microbial abundance profile, we simulate the model evolution with varying values of $m$. We run the independent evolution from time 0 to 500, switch to the coupled evolution between the two hosts from time 500 to 1000, and then remove the social interaction from time 1000 to time 1500. We randomly generate initial conditions on $U(0,1)$.

\begin{table}[h!]
\centering
\caption{Parameter Values and Initial Values}
\label{table10}
\begin{tabular}{|c|c|c|c|} 
 \hline
  strain & growth rate & carrying capacity & initial abundance \\
  \hline
  1 & 0.257104811123636 & 0.814904911789609 & 0.173771583224603 \\
  \hline
  2 & 0.295103582097683 & 1.19031910121444 & 0.651401064738548 \\
  \hline
  3 & 0.21554360521178 & 1.00893596320524 & 0.498696405983114 \\
  \hline
  4 & 0.294790614837217 & 1.16385199143913 & 0.284510688567093 \\
  \hline
  5 & 0.328856043925703 & 0.953299168536006 & 0.830559738735304 \\
  \hline
  6 & 0.272705794414955 & 1.15378089476191 & 0.81835988248334 \\
  \hline
  7 & 0.28957574461367 & 0.902007043732461 & 0.938169403753504 \\
  \hline
  8 & 0.18346120116613 & 1.16361993677712 & 0.000326096340118931 \\
  \hline
  9 & 0.351950926839943 & 1.1578241779005 & 0.640389224633762 \\
  \hline
  10 & 0.228050486871722 & 0.959406864396231 & 0.00735569277648584 \\
  \hline
  11 & 0.289486713504899 & 1.0500080007344 & 0.106421106692614 \\
  \hline
  12 & 0.35004009445623 & 1.02703891447434 & 0.106794413254953 \\
  \hline
  13 & 0.181055639084368 & 1.157804754178 & 0.3671090438542 \\
  \hline
  14 & 0.220240220851683 & 0.885666321547158 & 0.239607700586711 \\
  \hline
  15 & 0.266276377724096 & 0.801543663544288 & 0.346140201697452 \\
  \hline
  16 & 0.233159607049786 & 1.15223258441268 & 0.249619835400228 \\
  \hline
  17 & 0.127115236608583 & 0.894048654016205 & 0.387064314143913 \\
  \hline
  18 & 0.323314392996122 & 0.897945848938831 & 0.421038370433554 \\
  \hline
  19 & 0.109784634790871 & 1.0563668140559 & 0.640077289145426 \\
  \hline
  20 & 0.228922763029906 & 0.921809792483387 & 0.787552972956609 \\
 \hline
\end{tabular}
\end{table}

\begin{figure}[h!]
    \centering
    \begin{subfigure}[b]{0.45\textwidth}
        \includegraphics[width=\textwidth]{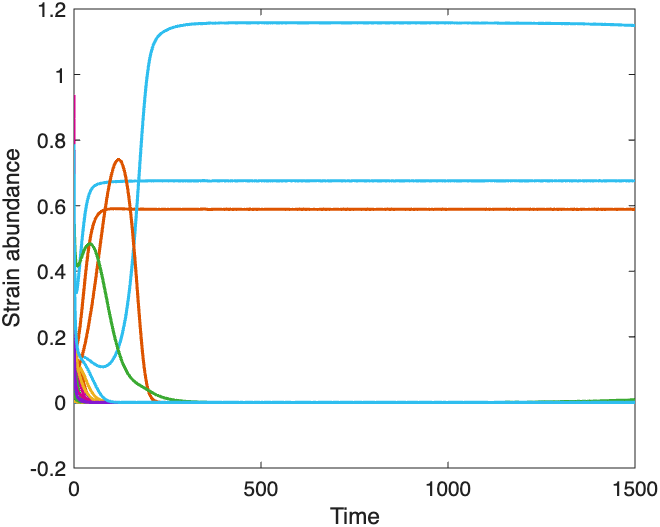}
        \caption{$m=0$}
    \end{subfigure}
    \hfill
    \begin{subfigure}[b]{0.45\textwidth}
        \includegraphics[width=\textwidth]{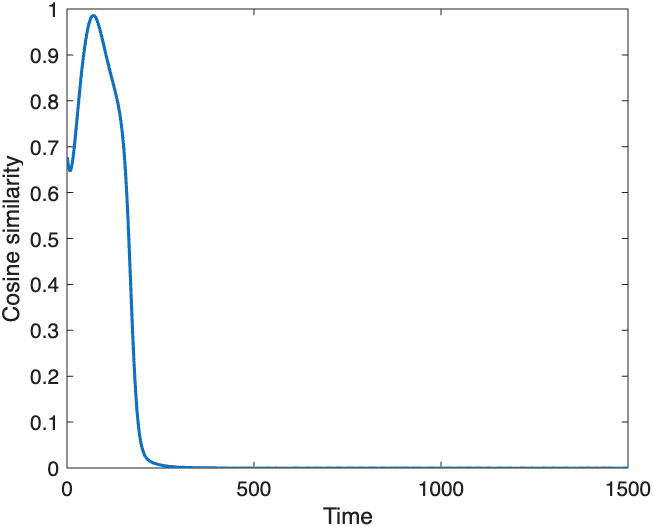}
        \caption{$m=0$}
    \end{subfigure}\\
    \begin{subfigure}[b]{0.45\textwidth}
        \includegraphics[width=\textwidth]{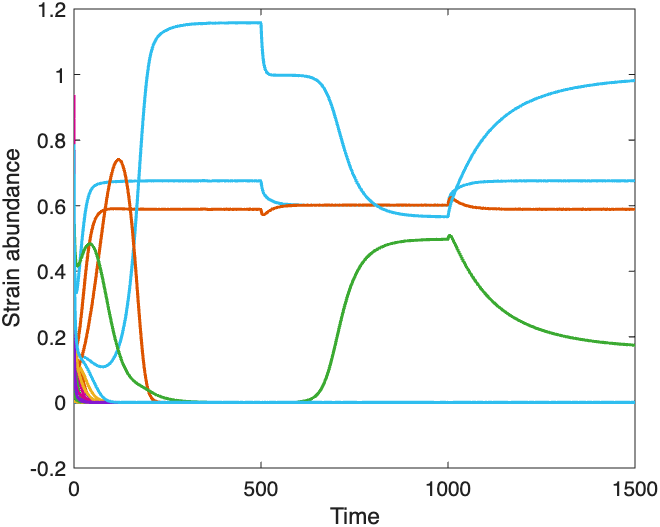}
        \caption{$m=0.025$}
    \end{subfigure}
    \hfill
    \begin{subfigure}[b]{0.45\textwidth}
        \includegraphics[width=\textwidth]{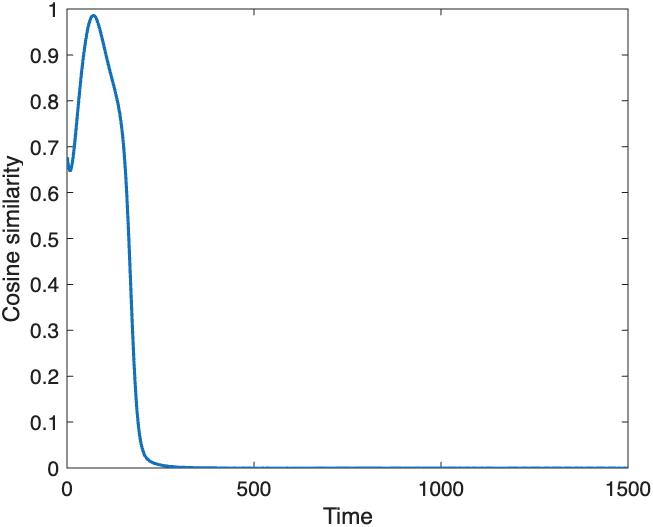}
        \caption{$m=0.025$}
    \end{subfigure}\\
    \begin{subfigure}[b]{0.45\textwidth}
        \includegraphics[width=\textwidth]{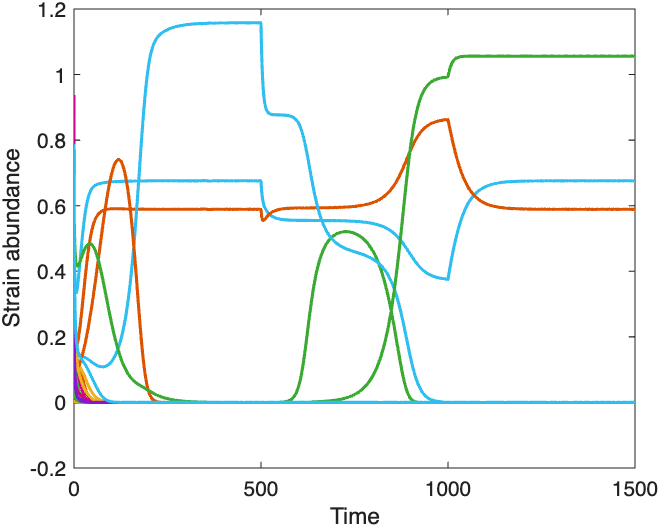}
        \caption{$m=0.05$}
    \end{subfigure}
    \hfill
    \begin{subfigure}[b]{0.45\textwidth}
        \includegraphics[width=\textwidth]{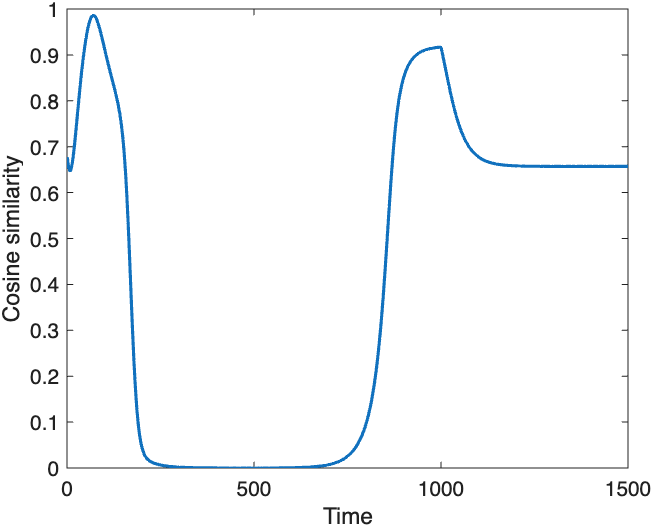}
        \caption{$m=0.05$}
    \end{subfigure}
    \caption{10-strain microbial dynamics in host 1 and host 2, part 1}
    \label{10-strain-1}
\end{figure}

\begin{figure}[h!]
    \centering
    \begin{subfigure}[b]{0.45\textwidth}
        \includegraphics[width=\textwidth]{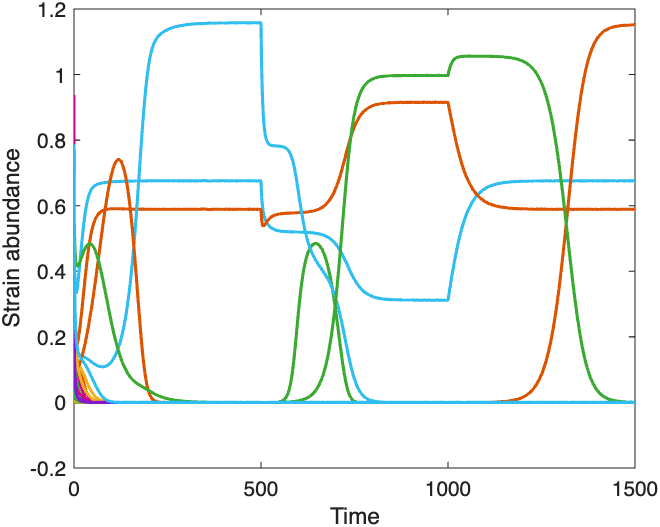}
        \caption{$m=0.075$}
    \end{subfigure}
    \hfill
    \begin{subfigure}[b]{0.45\textwidth}
        \includegraphics[width=\textwidth]{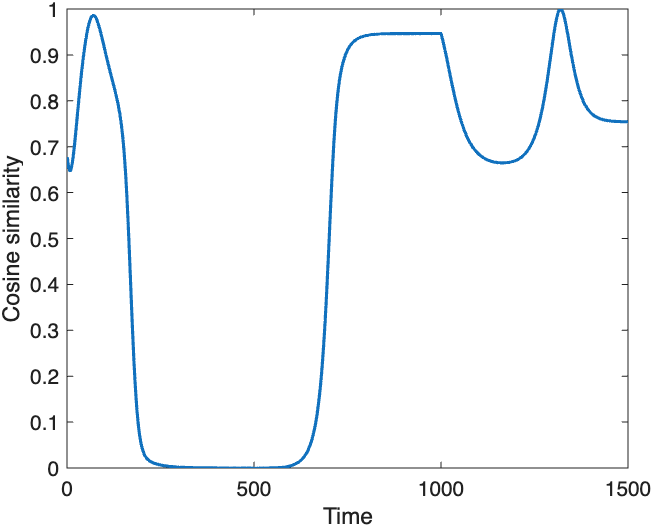}
        \caption{$m=0.075$}
    \end{subfigure}\\
    \begin{subfigure}[b]{0.45\textwidth}
        \includegraphics[width=\textwidth]{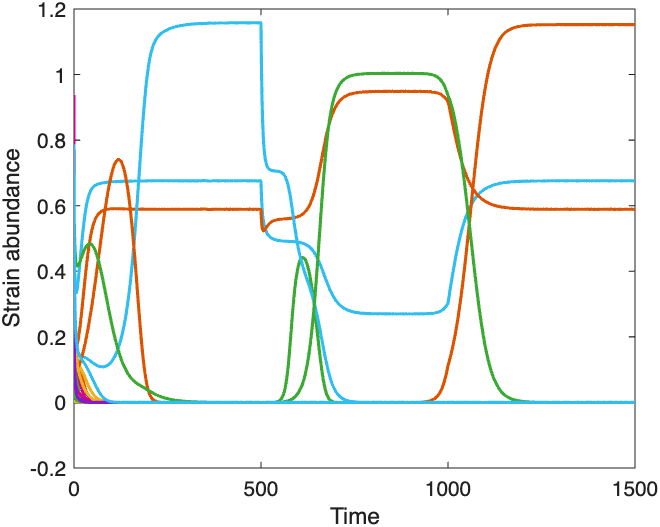}
        \caption{$m=0.1$}
    \end{subfigure}
    \hfill
    \begin{subfigure}[b]{0.45\textwidth}
        \includegraphics[width=\textwidth]{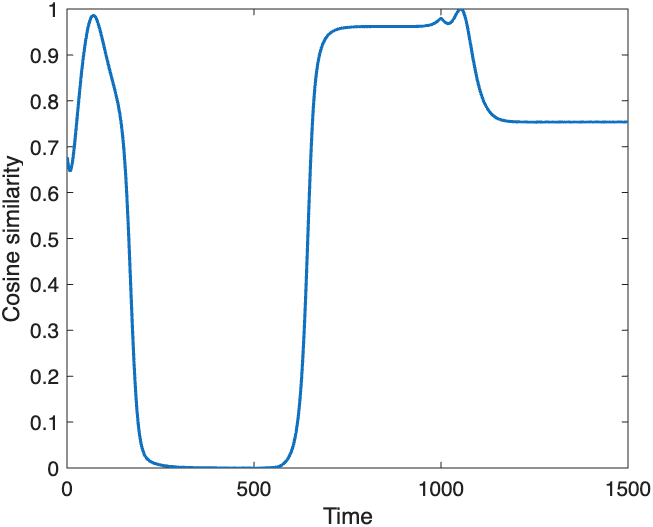}
        \caption{$m=0.1$}
    \end{subfigure}
    \caption{10-strain microbial dynamics in host 1 and host 2, part 2}
    \label{10-strain-2}
\end{figure}

Let strains 1 through 10 be the 10 strains of microbes in host 1 and strains 11 through 20 be the same corresponding strains in host 2. We have run multiple experiments with different randomly generated parameter values and initial conditions and list a specific case here. The parameter values and initial microbial abundance profile used are listed in table \ref{table10} and the competition matrix $B_1$ and $B_2$ are listed in appendix \ref{matrices-num}. We can see the evolution process in Figures \ref{10-strain-1} and \ref{10-strain-2}. There are 20 variables in total and it's hard to tell which strains are present from the plots, so we list the simulation results at times 500, 1000, and 1500 in appendix \ref{matrices-num} as well.

From Figure \ref{10-strain-1}(a) we can see the dynamics stabilizes around time 300 and from \ref{10-strain-1}(c), \ref{10-strain-1}(e), \ref{10-strain-2}(a) and \ref{10-strain-2}(c) we can see the social interaction changes the microbial dynamics. We plot the cosine similarity between the abundance profiles $P_1=\langle N_1^{(1)},N_2^{(1)},\dots,N_{10}^{(1)}\rangle$ of host 1 and $P_2=\langle N_1^{(2)},N_2^{(2)},\dots,N_{10}^{(2)}\rangle$ of host 2 to get a clearer understanding.

We can see $S_c(P_1,P_2)=\dfrac{P_1\cdot P_2}{\|P_1\|\|P_2\|}$ decreases to 0 in \ref{10-strain-1}(b) around time 300. In \ref{10-strain-1}(d) when $m=0.025$ the cosine similarity also decreases to 0 despite the microbial dynamics still evolving. In \ref{10-strain-1}(f) something interesting happens. At time 500 host 1 and host 2 start having social interactions but the cosine similarity stays at 0 until around time 700 and increases all the way to about 0.9 before the social interaction is removed at time 1000. And after time 1000, the cosine similarity decreases and stabilizes around 0.65. In this case, the social interaction has a lasting effect on the individual microbial profiles. In \ref{10-strain-2}(b) we see a similar trend in the cosine similarity. It increases after social interaction starts. Since $m$ is bigger now, the cosine similarity increases faster to the peak value and stays there before social interaction is removed. Then something very interesting happens again, the cosine similarity decreases at first and then goes back up to 1 before falling back down to about 0.75. Finally in \ref{10-strain-2}(d) when $m=0.1$ the cosine similarity drops back down to around 0.75 as well with a much less dramatic spike in the middle. One can check when $m>0.1$, we get a similar cosine similarity plot to \ref{10-strain-2}(d), just smoother. In addition to the plots, we list the strain abundance profiles for both hosts at times $t=500,t=1000,t=1500$ for the five different values of $m$ in appendix \ref{matrices-num}.

For the specific initial conditions and parameter values in this case, when $m\ge0.1$ we can numerically verify that the social interaction has a lasting effect on individual microbial profiles and the cosine similarity stabilizes around 0.75 in the long run. Technically we already see lasting effect when $m\ge0.05$ but as we can see when $m=0.075$ the cosine similarity goes through a transition phase from stabilizing around 0.65 in the long run to stabilizing around 0.75 in the long run and displays some oscillation in the process.

\subsection{A small social-network}
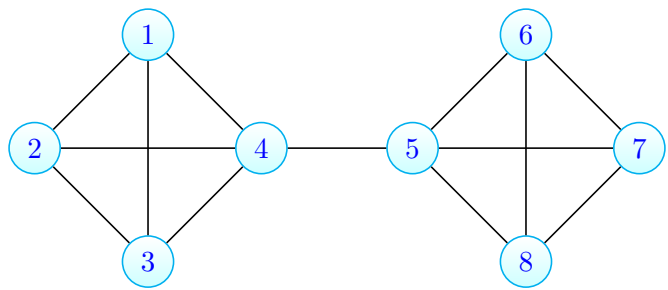
\begin {figure}[h]
\centering
\begin{tikzpicture}[-latex, auto, node distance = 0.3 cm and 0.3 cm, on grid, semithick, state/.style ={circle, top color =white, bottom color = processblue!20, draw, processblue, text=blue, minimum width =0.3 cm}]
\node[state] (A) at (0, 0) {$1$};
\node[state] (B) at (-1.5,-1.5) {$2$};
\node[state] (C) at (0,-3) {$3$};
\node[state] (D) at (1.5,-1.5) {$4$};
\node[state] (E) at (3.5,-1.5) {$5$};
\node[state] (F) at (5,0) {$6$};
\node[state] (G) at (6.5,-1.5) {$7$};
\node[state] (H) at (5,-3) {$8$};
\path[-] (A) edge (B) edge (C) edge (D)
(B) edge (C) edge (D)
(C) edge (D)
(D) edge (E)
(E) edge (F) edge (G) edge (H)
(F) edge (G) edge (H)
(G) edge (H);
\end{tikzpicture}
\caption{A small social network}\label{social}
\end{figure}

Now we look at a small social network with two groups of 4 hosts, each completely connected, and the two groups are connected through one member from each group, as seen in figure \ref{social}.

We consider the same ten strains of microbes in each host with growth rates randomly generated on $U(0.1,0.4)$, carrying capacities generated on $U(0.8,1.2)$, and the competition between different strains generated on $U(0.8,1.6)$. We run similar simulations as the 2 hosts 10 strain case and let the system evolve independently, i.e. no social interactions from time 0 to 500, then we add social interactions with strength $mA$, where $$A=\begin{pmatrix}
    0 & 1 & 1 & 1 & 0 & 0 & 0 & 0 \\
    1 & 0 & 1 & 1 & 0 & 0 & 0 & 0 \\
    1 & 1 & 0 & 1 & 0 & 0 & 0 & 0 \\
    1 & 1 & 1 & 0 & 1 & 0 & 0 & 0 \\
    0 & 0 & 0 & 1 & 0 & 1 & 1 & 1 \\
    0 & 0 & 0 & 0 & 1 & 0 & 1 & 1 \\
    0 & 0 & 0 & 0 & 1 & 1 & 0 & 1 \\
    0 & 0 & 0 & 0 & 1 & 1 & 1 & 0
\end{pmatrix}$$ is the adjacency matrix of our symmetric social network of size 8. We run the evolution with social interactions from time 500 to time 1000, then remove the social interactions after time 1000 and run the simulation until time 1500. We run these simulations with different $m$ values and present the results below.

\begin{figure}[h!]
    \centering
    \begin{subfigure}[b]{0.45\textwidth}
        \includegraphics[width=\textwidth]{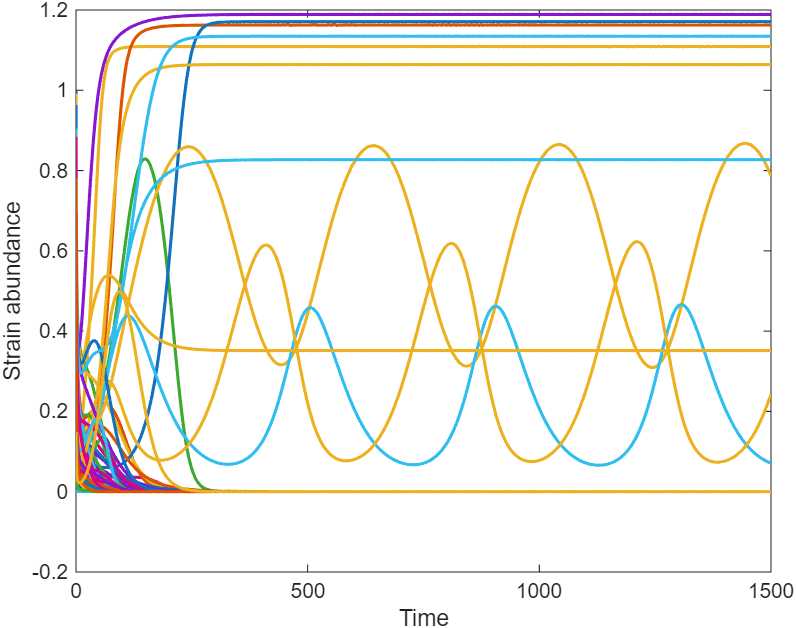}
        \caption{$m=0$}
    \end{subfigure}
    \hfill
    \begin{subfigure}[b]{0.45\textwidth}
        \includegraphics[width=\textwidth]{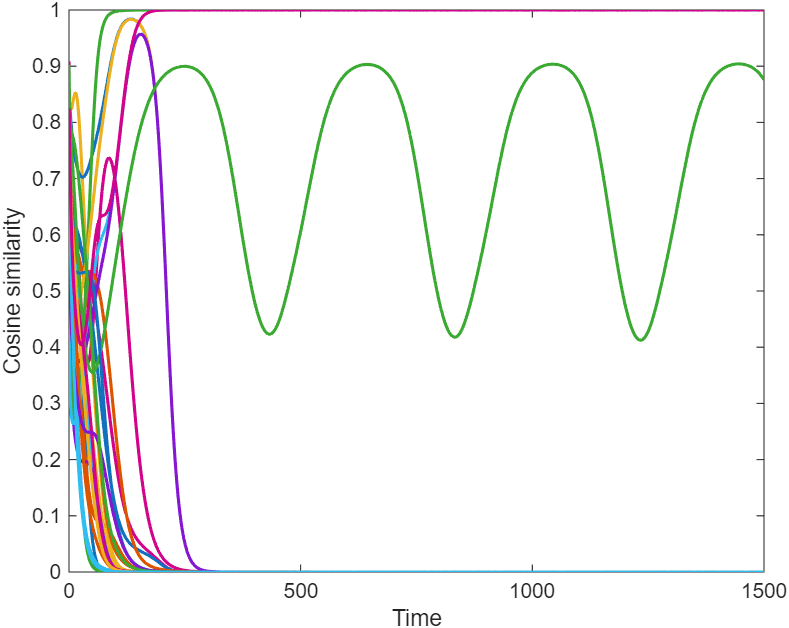}
        \caption{$m=0$}
    \end{subfigure}\\
    \begin{subfigure}[b]{0.45\textwidth}
        \includegraphics[width=\textwidth]{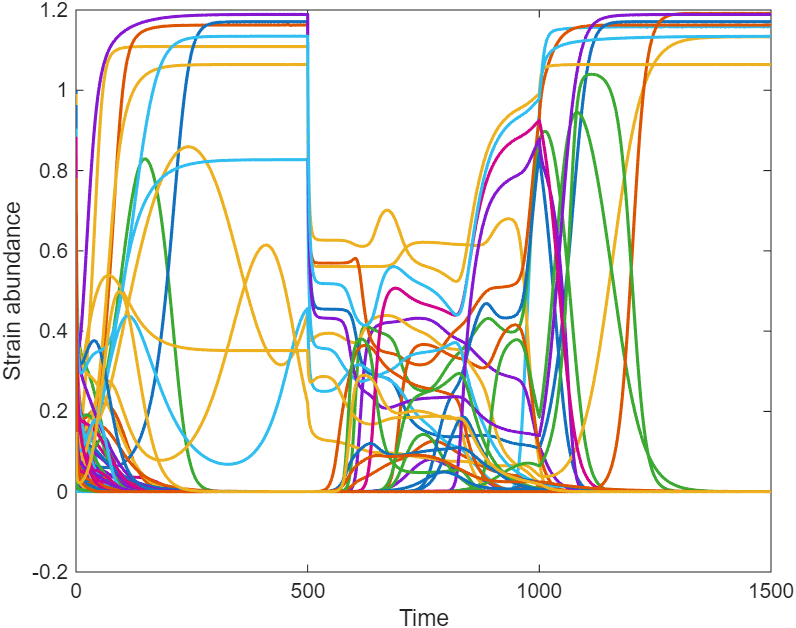}
        \caption{$m=0.1$}
    \end{subfigure}
    \hfill
    \begin{subfigure}[b]{0.45\textwidth}
        \includegraphics[width=\textwidth]{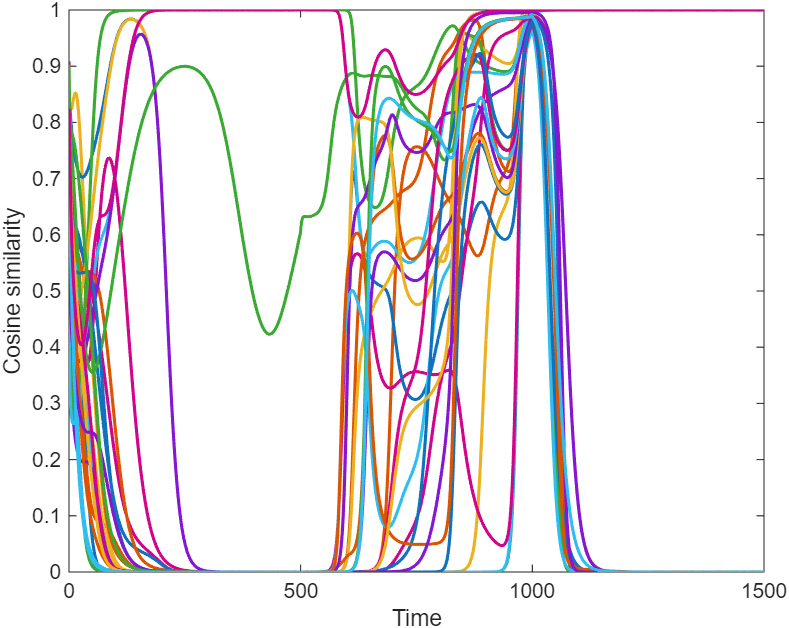}
        \caption{$m=0.1$}
    \end{subfigure}
    \caption{Microbial evolution in a small social network}
    \label{10-strain-8}
\end{figure}

Surprisingly, with the specific instance of parameters and initial conditions used (available upon request, not listed here since there are 80 strains and many parameters), we have a periodic solution when there is no social interaction at all, see Figure \ref{10-strain-8}(a). In this case the pairwise cosine similarity also has a periodic solution, see \ref{10-strain-8}(b). We run the experiment with values of $m$ set at 0, 0.025, 0.05, 0.075 and 0.1. We only show the results for $m=0.1$ here in figure \ref{10-strain-8}(c) and (d). As one can see, the dynamics is quite complicated. The social interaction at time 500 has an effect on the dynamics immediately and alters the dynamics after the social interaction is removed as well.

\begin{figure}[h]
    \centering
    \begin{subfigure}[b]{0.45\textwidth}
        \includegraphics[width=\textwidth]{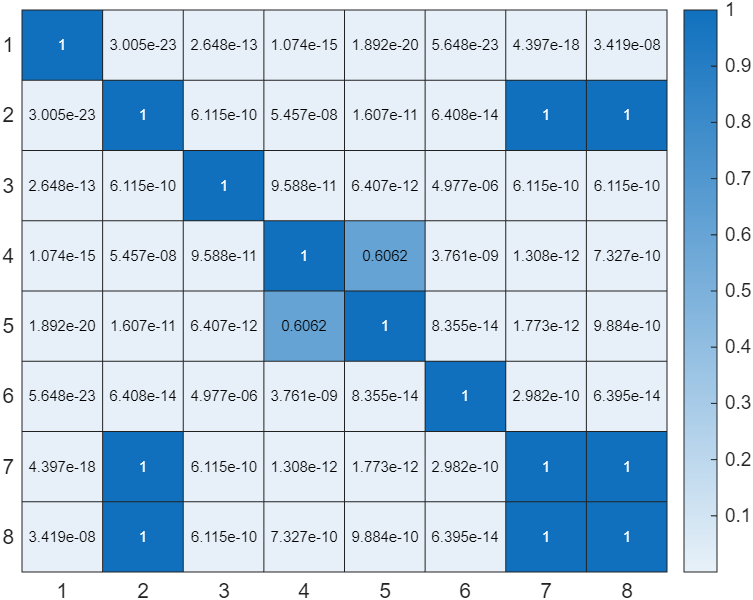}
        \caption{$t=500$}
    \end{subfigure}
    \hfill
    \begin{subfigure}[b]{0.45\textwidth}
        \includegraphics[width=\textwidth]{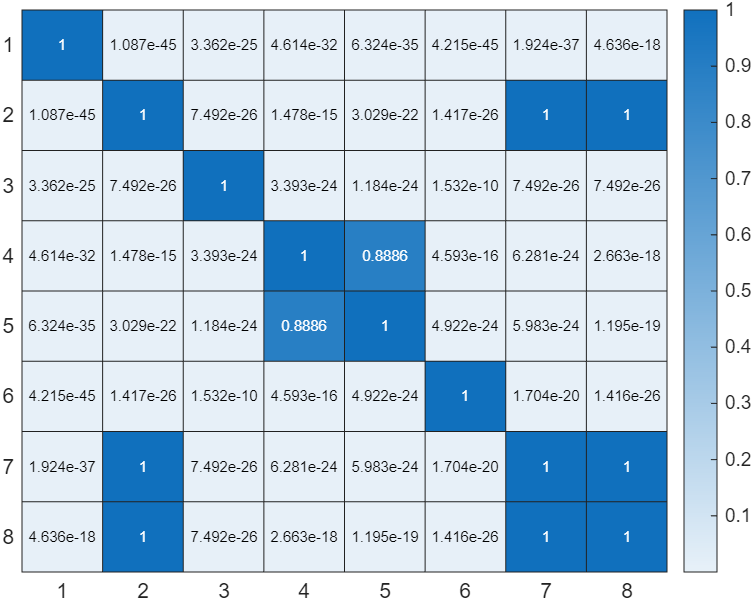}
        \caption{$t=1000$}
    \end{subfigure}\\
    \begin{subfigure}[b]{0.45\textwidth}
        \includegraphics[width=\textwidth]{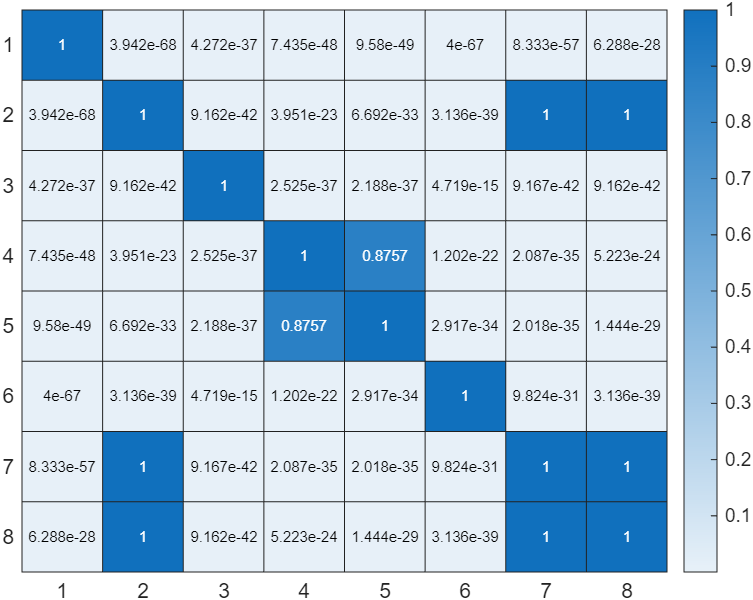}
        \caption{$t=1500$}
    \end{subfigure}
    \caption{Microbial evolution without social interaction, $m=0$}
    \label{heatmap-8}
\end{figure}

Since there are 80 strains in total and 28 pair-wise cosine similarity curves, it's difficult to identify the specific strain or host in figure \ref{10-strain-8}. To better visualize the experiment result, we plot heatmaps of the pair-wise cosine similarity at time 500, time 1000, and time 1500 for $m=0$ in Figure \ref{heatmap-8}. Without social interaction, hosts 2, 7, and 8 have the same microbial profile and cosine similarity 1. They all have only strain 6 (I know because I printed out the abundance of all 80 strains at times 500, 1000, and 1500. All original Matlab programs are available upon request). Hosts 4 and 5 have a high cosine similarity that is oscillating. Host 4 has strains 1, 4 and 8 present while host 5 has strains 1 and 5. And the abundance of strains 4 and 8 in host 4 are oscillating over time.

\begin{figure}[h]
    \centering
    \begin{subfigure}[b]{0.45\textwidth}
        \includegraphics[width=\textwidth]{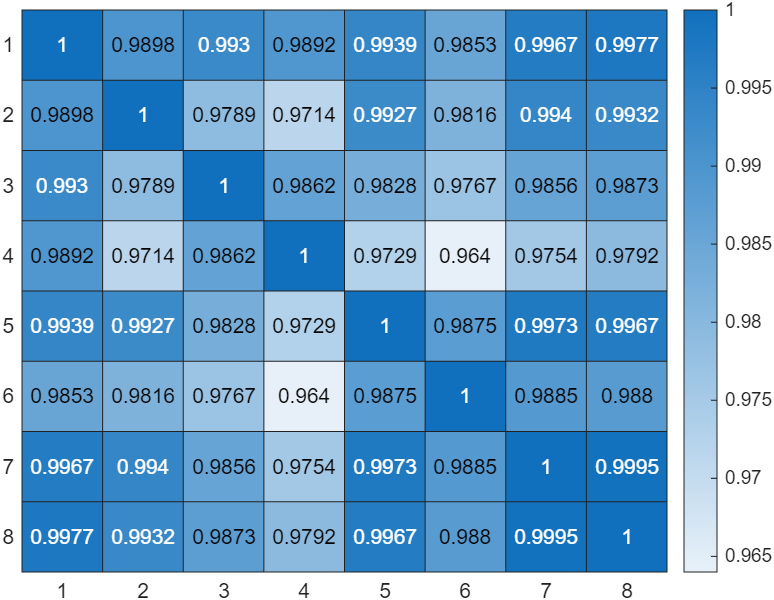}
        \caption{$m=0.1, t=1000$}
    \end{subfigure}
    \hfill
    \begin{subfigure}[b]{0.45\textwidth}
        \includegraphics[width=\textwidth]{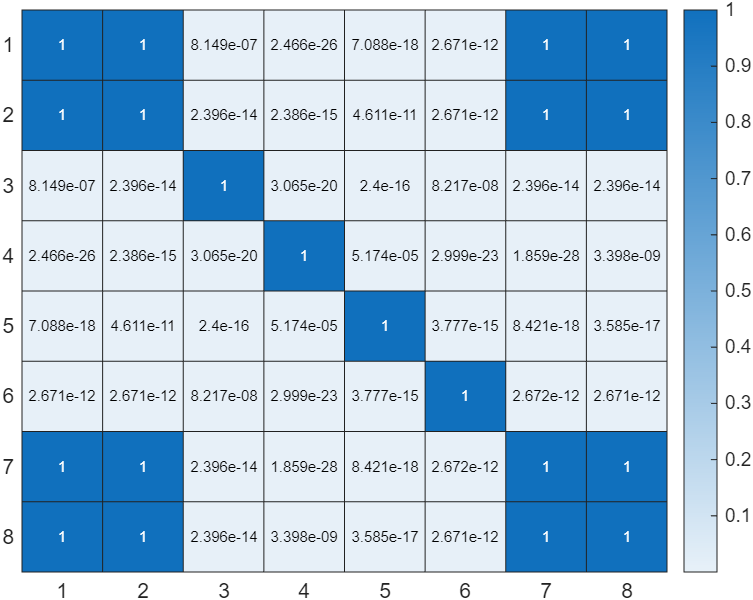}
        \caption{$m=0.1, t=1500$}
    \end{subfigure}
    \caption{Microbial evolution with social interaction in a small social network}
    \label{heatmap-8-1}
\end{figure}

From Figure \ref{heatmap-8-1} we can see when $m=0.1$, the pairwise cosine similarity generally increases after social interaction at time 1000. By time 1500, after the social interaction is removed, we can see hosts 1, 2, 7 and 8 are in sync with only strain 6 present. The oscillation in host 4 disappears and only strain 8 is present and host 5 only has strain 4 present.

\begin{figure}[h!]
    \centering
    \begin{subfigure}[b]{0.45\textwidth}
        \includegraphics[width=\textwidth]{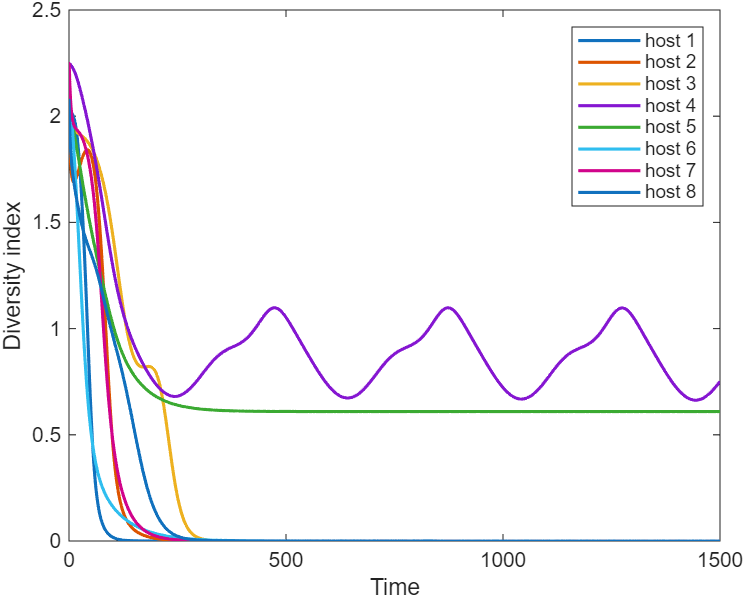}
        \caption{$m=0$}
    \end{subfigure}
    \hfill
    \begin{subfigure}[b]{0.45\textwidth}
        \includegraphics[width=\textwidth]{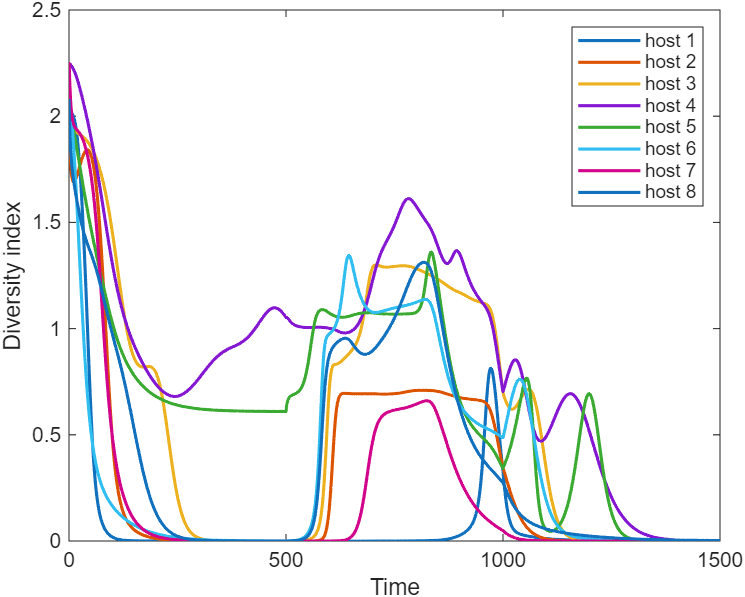}
        \caption{$m=0.1$}
    \end{subfigure}
    \caption{Shannon index for the hosts over time}
    \label{shannon}
\end{figure}

Finally, we plot the Shannon index
$$H=\sum_{i=1}^{10}-p_i\ln(p_i)$$ for each host over time. Here $p_i$ is the ratio between the abundance of strain $i$ and the sum of all strain abundances. Notice that $-p_i\ln(p_i)=0$ when $p_i=0$ or 1. The Shannon index reflects in-host microbial diversity. Figure \ref{shannon} shows that without social interactions, all hosts end up with only one strain of microbe except host 4 and 5 and the diversity index of host 4 oscillates over time. With social interaction at strength $m=0.1$ from time 500 to time 1000, we can see every host's diversity index increases during the social interaction. However after the social interaction is removed every host's diversity decreases to 0, meaning each host only has one strain present. Judging by figure \ref{10-strain-8} the dynamics reached equilibrium by time 1500, so the social interaction has a lasting effect in this specific case. It pushes the system out of a limit cycle and into the attraction basin of a fixed point where each host has only one strain present.

\section{Discussions and future directions}
\label{sec:discussions}
We started with a simple 2-host 2 strain model here to examine the effect of social interaction on the microbial profile of each individual host. From the simple example we get the intuition that strong in-host competition promotes multiple local sinks in the individual dynamics instead of one global sink. A global sink generally indicates the robustness of the system; it always goes back to the global sink, no matter the perturbation. Given the delicate and complex nature of the human gut microbiome we have a reason to believe that multiple local sinks instead of one global sink matches the reality of microbial dynamics. Our analysis and numerical experiments with more strains in each host, and more hosts verify that with a strong in-host competition between the different strains, social interaction typically has a lasting impact on the individual microbial dynamics. The individual equilibrium reached without social interaction would be altered after a limited amount of time of social interaction. The social interaction pushes the individual dynamics out of the attraction basin of one local sink to another. We can also see from the plots that the dynamics is quite rich, we see oscillations in transitions and also as steady states.

In the future we hope to expand our modeling and take real data into consideration.

\appendix
\section{Verification of the edges in the invasion graph in figure \ref{invasion-inter-2}}
\label{invasion-verf}
Let's recall that in an invasion graph, there is an edge from node $S$ to node $T$ if 
\begin{enumerate}
    \item[i)] $S\ne T$,
    \item[ii)] $g(S)$ has positive signs on all components in $T\setminus S$,
    \item[iii)] $g(T)$ has negative signs on all components in $S\setminus T$.
\end{enumerate}
We divide the invasion graph into 4 levels.
\begin{enumerate}
    \item Level 0
    \begin{itemize}
        \item $\emptyset$ \\
        $g(\emptyset)=(+,+,+,+)$. $\forall T\neq\emptyset$, $g(\emptyset)$ is positive on $T\setminus\emptyset=T$; and $\emptyset\setminus T=\emptyset$. So there is an edge from $\emptyset$ to every other node.
    \end{itemize}
    \item Level 1
    \begin{itemize}
        \item $x_1$ \\
        $g(x_1)=(0,+,+,+)$. $\forall T\supsetneq\{x_1\}$, $g(x_1)$ is positive on $T\setminus\{x_1\}$ and $\{x_1\}\setminus T=\emptyset$. There is an edge from $x_1$ to $x_1y_1,x_1x_2,x_1y_2,x_1y_1x_2,x_1y_1y_2,$ and $x_1y_1x_2y_2$. $\forall T\not\ni x_1$, we have $\{x_1\}\setminus T=\{x_1\}$; we need a node with $g(T)=(-,\dots)$. No node has a negative sign in the first coordinate of the growth rates.
        \item $y_1$ \\
        $g(y_1)=(+,0,+,+)$. A parallel argument to the one above concludes there is an edge from $y_1$ to $x_1y_1,y_1x_2,y_1y_2,x_1y_1x_2,x_1y_1y_2$, and $x_1y_1x_2y_2$. And no node has a negative sign in the second coordinate of the growth rates.
        \item $x_2$ \\
        $g(x_2)=(+,+,0,-)$. $\forall T\supsetneq\{x_2\}$, we have $\{x_2\}\setminus T=\emptyset$ and just need $g(x_2)$ to be positive on $T\setminus\{x_2\}$; therefore $T\setminus\{x_2\}\subset\{x_1,y_1\}$. There is an edge from $x_2$ to $x_1x_2,y_1x_2,$ and $x_1y_1x_2$. $\forall T\not\ni x_2$, we have $\{x_2\}\setminus T=\{x_2\}$ and need a negative sign in the third coordinate of the growth rates. Nodes $y_2,y_1y_2,$ and $x_1y_1y_2$ satisfy this condition. But all three nodes satisfy $T\setminus \{x_2\}\ni y_2$ and $g(x_2)$ has a negative sign in the fourth coordinate.
        \item $y_2$ \\
        $g(y_2)=(+,+,-,0)$. $\forall T\supsetneq\{y_2\}$, we need $g(y_2)$ to be positive on $T\setminus\{y_2\}$, so $T\setminus\{y_2\}\subset\{x_1,y_1\}$. There is an edge from $y_2$ to $x_1y_2,y_1y_2$, and $x_1y_1y_2$. $\forall T\not\ni y_2$, $\{y_2\}\setminus T=\{y_2\}$, we need a node with a negative sign in the fourth coordinate. $x_2,x_1x_2,$ and $x_1y_2x_2$ satisfy the condition however they all have $x_2\in T\setminus\{y_2\}$ and $g(y_2)$ is negative at the third coordinate. 
    \end{itemize}
    \item Level 2
    \begin{itemize}
        \item $x_1y_1$ \\
        $g(x_1y_1)=(0,0,+,+)$. $\forall T\supsetneq\{x_1,y_1\}$, we need $g(x_1y_1)$ positive on $T\setminus\{x_1,y_1\}$. So there is an edge from $x_1y_1$ to $x_1y_1x_2$, $x_1y_1y_2$, and $x_1y_1x_2y_2$. For the rest of the nodes, we need $g(T)$ to have negative signs on $\{x_1,y_1\}\setminus T$. But no nodes have negative signs on $x_1$ or $y_1$.
        \item $x_1x_2$ \\
        $g(x_1x_2)=(0,+,0,-)$. If $T\supsetneq\{x_1,x_2\}$, then $T\setminus\{x_1,x_2\}=\{y_1\}$ and there is an edge from $x_1x_2$ to $x_1y_1x_2$. For other nodes, we need $g(T)$ to have negative signs on $\{x_1,x_2\}\setminus T$. No nodes have negative signs in the first coordinate and only $y_1y_2$ and $x_1y_1y_2$ have negative signs in the third coordinate. They both have $y_2$ which has a negative sign in $g(x_1x_2)$.
        \item $x_1y_2$ \\
        $g(x_1y_2)=(0,+,+,0)$. $\forall T\supsetneq\{x_1,y_2\}$, we have $g(x_1y_2)$ is positive on $T\setminus\{x_1,y_2\}$. There is an edge from $x_1y_2$ to $x_1y_1y_2$ and $x_1y_1x_2y_2$. For other nodes, we need $g(T)$ to have negative signs on $\{x_1,y_2\}\setminus T$. No node has a negative sign in the first coordinate and nodes $x_2,x_1x_2$, and $x_1y_1x_2$ have a negative sign in the fourth coordinate. Only the latter two nodes satisfy condition iii) so there is an edge from $x_1y_2$ to $x_1x_2$ and $x_1y_1x_2$.
        \item $y_1x_2$ \\
        $g(y_1x_2)=(+,0,0,+)$. For nodes $T\supsetneq\{y_1,x_2\}$ we have an edge from $y_1x_2$ to $x_1y_1x_2$ and $x_1y_1x_2y_2$. For other nodes, we need $g(T)$ to be negative in the second or third coordinate. No node has a negative second coordinate. Nodes with negative third coordinate are $y_2,y_1y_2$, and $x_1y_1y_2$. The latter two satisfy condition iii) and we have an edge from $y_1x_2$ to $y_1y_2$ and $x_1y_1y_2$.
        \item $y_1y_2$ \\
        $g(y_1y_2)=(+,0,-,0)$. For nodes $T\supsetneq\{y_1,y_2\}$, to satisfy condition ii), $T$ can only be $x_1y_1y_2$. There is an edge from $y_1y_2$ to $x_1y_1y_2$. For other nodes we need condition iii), $g(T)$ has negative signs on $\{y_1,y_2\}\setminus T$. No node has a negative sign in the second coordinate, only nodes $x_2,x_1x_2,x_1y_1x_2$ have a negative sign in the fourth coordinate. But none of them satisfy condition ii).
        \item $x_2y_2$ \\
        $g(x_2y_2)=(+,+,0,0)$. Any node $T\supsetneq\{x_2,y_2\}$ satisfy condition ii). There is an edge from node $x_2y_2$ to $x_1y_1x_2y_2$. For other nodes, we need condition iii), $g(T)$ has negative signs on $\{x_2,y_2\}\setminus T$. Nodes $x_2,y_2,x_1x_2,y_1y_2,x_1y_1x_2,x_1y_1y_2$ have negative signs in either the third or fourth coordinate. They all satisfy both conditions ii) and iii), so there is an edge from node $x_2y_2$ to nodes $x_2,y_2,x_1x_2,y_1y_2,x_1y_1x_2,x_1y_1y_2$.
    \end{itemize}
    \item Level 3
    \begin{itemize}
        \item $x_1y_1x_2$ \\
        $g(x_1y_1x_2)=(0,0,0,-)$. For condition iii), we need $T\supset\{x_1,y_1\}$ since no node has negative signs in the first two coordinates. We only have nodes $x_1y_1y_2$ and $x_1y_1x_2y_2$ to consider. Neither satisfies condition ii) so there is no outgoing edge from node $x_1y_1x_2$.
        \item $x_1y_1y_2$ \\
        $g(x_1y_1y_2)=(0,0,-,0)$. Again to satisfy condition iii) we have to consider nodes $T\supset\{x_1,y_1\}$, i.e. nodes $x_1y_1x_2,x_1y_1x_2y_2$. Neither satisfies condition ii) so there is no outgoing edge from node $x_1y_1y_2$.
    \end{itemize}
    \item Level 4
    \begin{itemize}
        \item $x_1y_1x_2y_2$ \\
        $g(x_1y_1x_2y_2)=(0,0,0,0)$. For every other node we have $T\setminus\{x_1,y_1,x_2,y_2\}=\emptyset$ so condition ii) automatically holds. For condition iii), we need nodes with negative signs again, i.e. nodes $x_2,y_2,x_1x_2,y_1y_2,x_1y_1x_2,x_1y_1y_2$. Furthermore there cannot be positive signs, the only nodes left are $x_1y_1x_2$ and $x_1y_1y_2$. There is an edge from $x_1y_1x_2y_2$ to $x_1y_1x_2$ and $x_1y_1y_2$.
    \end{itemize}
\end{enumerate}

\section{Competition matrices and simulation results in the 2-host 10-strain case}
\label{matrices-num}
For the numerical experiment result plotted in Figures \ref{10-strain-1} and \ref{10-strain-2}, the competition matrices for host 1 and 2 are listed in tables \ref{table-b1} and \ref{table-b2} below. For $t=500$, all the runs have the same results since there is no social interaction from time 0 to time 500. So we list the strain abundance profile at time 500, 1000, and 1500 for $m=0$ in table \ref{2-10-1}, and the strain abundance profile for other values of $m$ at time 1000 in table \ref{2-10-2} and the results at time 1500 in table \ref{2-10-3}.
From table \ref{2-10-1} we can tell with no social interactions, host 1 stabilizes with only strains 6 and 9 and host 2 stabilizes with only strains 2 and 3. Tables \ref{2-10-2} and \ref{2-10-3} show that with social interaction, host 1 still has only strains 6 and 9 but host 2 transitions from strains 2 and 3 to 6 and 9 and finally to only strain 6.

\begin{table}[h!]
\caption{Host 1 competition matrix}
\label{table-b1}
\begin{tabular}{|c|c|c|c|c|} 
  \hline
  column 1 & column 2 & column 3 & column 4 & column 5 \\
  \hline
  1 & 1.51085321771245 & 1.2978439877721 & 1.42687231054242 & 0.900368285312024 \\
  \hline
  1.50695096875781 & 1 & 1.3727243266339 & 1.34164965015808 & 1.40179444099712 \\
  \hline
  1.55629836638837 & 0.987828420243464 & 1 & 0.91984938401512 & 1.46164152756493 \\
  \hline
  1.1126362707237 & 1.47172578217939 & 1.12981852683926 & 1 & 1.4251442615282 \\
  \hline
  1.44105605176933 & 1.19643202378075 & 1.0897645707313 & 0.903210434498196 & 1 \\
  \hline
  0.925690630381974 & 0.921893093130324 & 1.42511361760202 & 1.55675702993257 & 1.14291307248318 \\
  \hline
  1.30013231742392 & 0.984614524647597 & 0.908389721041585 & 1.50912947939429 & 0.811565397693033 \\
  \hline
  1.35918828220772 & 1.32636316566776 & 1.52165561163974 & 1.21200369034994 & 1.06022839097576 \\
  \hline
  0.868695146496348 & 1.25035870499206 & 1.03170886870251 & 1.34352699674531 & 0.907761622901096 \\
  \hline
  1.22494365823075 & 1.03346299011908 & 1.19964634088784 & 1.58143468626691 & 1.16041396966722 \\
  \hline
  column 6 & column 7 & column 8 & column 9 & column 10 \\
  \hline
  1.25781994414286 & 1.06792366350215 & 1.34415517155844 & 1.15831171487017 & 1.32763831052924 \\
  \hline
  1.43361862558654 & 0.98294649379894 & 1.53393899875795 & 1.32099893057355 & 1.03581837621415 \\
  \hline
  1.13578928798244 & 1.45792144611659 & 1.00535333672093 & 0.935601496491956 & 1.56029472121192 \\
  \hline
  1.22602923547276 & 1.07858766319474 & 1.50849450754322 & 1.22515939308766 & 1.35542899805244 \\
  \hline
  1.54056351362241 & 0.932376476294587 & 1.53603423209455 & 1.30704068655517 & 0.965445221616733 \\
  \hline
  1 & 0.822506854406606 & 1.04005075202935 & 0.811276803346154 & 1.24380986076539 \\
  \hline
  1.23586671877499 & 1 & 0.858712757355034 & 1.1762970681848 & 1.50342278760953 \\
  \hline
  1.52089916013629 & 1.34423256936448 & 1 & 1.50906077693448 & 1.24628615257584 \\
  \hline
  0.841461367123206 & 1.48844969177973 & 0.867961772692011 & 1 & 1.40186724812113 \\
  \hline
  1.44688201494409 & 1.55127481228195 & 1.38301126367414 & 1.15403292132014 & 1 \\
  \hline
\end{tabular}
\end{table}

\begin{table}[h!]
\caption{Host 2 competition matrix}
\label{table-b2}
\begin{tabular}{|c|c|c|c|c|} 
  \hline
  column 1 & column 2 & column 3 & column 4 & column 5 \\
  \hline
  1 & 1.04091756168068 & 1.17516168396475 & 0.956044258144868 & 1.34403355650264 \\
  \hline
  0.904685346116759 & 1 & 0.869812027568055 & 1.36430342371673 & 1.21196144427756 \\
  \hline
  0.951323147081591 & 1.0065304187987 & 1 & 0.944438602932674 & 1.21765510489526 \\
  \hline
  0.922911864714045 & 1.38628349110124 & 1.34875576675108 & 1 & 0.882334863377167 \\
  \hline
  0.82312165253208 & 0.893408600833212 & 1.0138600440766 & 1.0369374586288 & 1 \\
  \hline
  0.807267918793451 & 1.39683325060608 & 1.57558685167396 & 1.17022551062748 & 1.0871750554996 \\
  \hline
  1.27716139357763 & 1.44783128567535 & 0.947019362132668 & 1.54018593334342 & 1.30019407902417 \\
  \hline
  1.28723945950705 & 1.39618694496933 & 1.03995273835592 & 0.972711313836821 & 1.11469321698024 \\
  \hline
  1.53513874560844 & 1.0697146512907 & 1.12894791255109 & 0.80080710280561 & 0.806129956400323 \\
  \hline
  1.3868595546985 & 1.26745998434907 & 0.989191349692337 & 1.52528499449347 & 1.23622794722568 \\
  \hline
  column 6 & column 7 & column 8 & column 9 & column 10 \\
  \hline
  1.20728829955727 & 1.3821579549539 & 1.23811419381862 & 1.36755623320316 & 1.04003297208718 \\
  \hline
  0.997423136257119 & 1.3207894377011 & 1.11610910067633 & 1.59409584742513 & 1.45083585911759 \\
  \hline
  0.836316219729312 & 1.33169201611993 & 1.1186175956201 & 1.54575589607408 & 0.861367899360652 \\
  \hline
  1.47338379580881 & 1.55102384041621 & 1.40107892090071 & 0.873783597111079 & 1.08357847635124 \\
  \hline
  0.83859993408450 & 1.22806501257422 & 1.21788027684366 & 1.56283197826357 & 0.905608699527181 \\
  \hline
  1 & 1.11875209472808 & 1.192346426921 & 0.930236527054001 & 0.926543704267675 \\
  \hline
  1.42673554435142 & 1 & 0.870943147633982 & 1.57636406981746 & 0.849717641293894 \\
  \hline
  1.5779200477698 & 1.15242757742115 & 1 & 1.2776053322143 & 1.36147478311386 \\
  \hline
  1.26917126459122 & 0.906299155822448 & 1.15804721822402 & 1 & 0.869185357663573 \\
  \hline
  1.42243507942529 & 1.15136299835343 & 1.31036871686537 & 0.856236129182204 & 1 \\
  \hline
\end{tabular}
\end{table}

\begin{table}[h]
    \centering
    \caption{Strain abundance profile for 2 hosts each with 10 strains, $m=0$}
    \label{2-10-1}
    \begin{tabular}{|c|c|c|c|}
    \hline
    strain & $t=500$ & $t=1000$ & $t=1500$ \\
    \hline
    1 & 2.22871058382802e-26 & 1.23913455482189e-53 & 8.86279343949447e-81 \\
    \hline
    2 & 1.28092077155892e-24 & 2.77285211712669e-53 & 6.5905584785526e-82 \\
    \hline
    3 & 5.13892918066967e-15 & 2.18183082052853e-29 & 9.2706987259824e-44 \\
    \hline
    4 & 2.93369172546466e-20 & 2.07831853403014e-41 & 1.49778180349233e-62 \\
    \hline
    5 & 1.26869927134853e-13 & 2.49873585780963e-18 & 7.10418374655657e-23 \\
    \hline
    6 & 0.67574774134966 & 0.675756546356759 & 0.675735410387507 \\
    \hline
    7 & 8.17411763841885e-27 & 1.99604917536623e-58 & 6.11573269864967e-90 \\
    \hline
    8 & 5.66743766846275e-26 & 2.89333548263813e-51 & 1.55934264455296e-76 \\
    \hline
    9 & 0.5890726538901 & 0.589081110869097 & 0.589056650997145 \\
    \hline
    10 & 3.26013339613057e-29 & 2.88194814927519e-61 & 2.96353485830336e-93 \\
    \hline
    11 & 9.93162062556937e-18 & 2.81624349803168e-36 & 8.29958143652294e-55 \\
    \hline
    12 & 9.91244468112135e-06 & 0.000297096940330678 & 0.00836344115389753 \\
    \hline
    13 & 1.15776317754052 & 1.15751652837445 & 1.14967508274116 \\
    \hline
    14 & 7.73441764558691e-24 & 5.29209840021473e-56 & 4.19136015622524e-88 \\
    \hline
    15 & 4.89971598876291e-23 & 3.32232734255503e-49 & 2.48071270949318e-75 \\
    \hline
    16 & 9.35648431347462e-15 & 4.89838746810574e-43 & 2.87636302594976e-71 \\
    \hline
    17 & 1.17667365374394e-10 & 6.60654980895425e-17 & 3.39353054626497e-23 \\
    \hline
    18 & 4.69539175449738e-27 & 1.11676061862024e-50 & 2.35562729030596e-74 \\
    \hline
    19 & 1.77007426391791e-05 & 3.88743059525688e-11 & 8.56500479310064e-17 \\
    \hline
    20 & 8.56783254680117e-14 & 7.61825570607756e-26 & 6.18989752950233e-38 \\
    \hline
    \end{tabular}
\end{table}

\begin{table}[h]
    \centering
    \caption{Strain abundance profile for 2 hosts each with 10 strains, $t=1000$}
    \label{2-10-2}
    \begin{tabular}{|c|c|c|c|c|}
    \hline
    strain & $m=0.025$ & $m=0.05$ & $m=0.075$ & $m=0.1$ \\
    \hline
    1 & 1.21252064308941e-61 & 3.03339979865282e-62 & 9.72416354567372e-64 & 8.2954337914018e-65 \\
    \hline
    2 & 3.34435791305472e-48 & 1.23289006794356e-44 & 8.41408739977226e-45 & 2.13540536043875e-44 \\
    \hline
    3 & 6.29523436868733e-22 & 7.81590359193665e-19 & 1.28421309536635e-18 & 4.34882713379456e-18 \\
    \hline
    4 & 2.14567172899838e-37 & 2.5859493300681e-35 & 2.45480715089529e-35 & 5.62211539401938e-35 \\
    \hline
    5 & 6.53576586537095e-35 & 5.86085748679764e-45 & 2.28764425863602e-52 & 1.62692833855941e-58 \\
    \hline
    6 & 0.601847022264451 & 0.375541729238047 & 0.31209255567032 & 0.302089745477119 \\
    \hline
    7 & 8.598319807518e-62 & 2.2606272298903e-60 & 8.592604949129e-61 & 7.29745362020828e-61 \\
    \hline
    8 & 1.64041163324042e-48 & 6.57692048840921e-47 & 6.61516571502909e-47 & 1.27092879855496e-46 \\
    \hline
    9 & 0.602050978075724 & 0.862508735456472 & 0.915333658941793 & 0.916534144306635 \\
    \hline
    10 & 8.67944967438523e-60 & 4.7087542613764e-57 & 1.08634744046198e-55 & 1.51185598047508e-54 \\
    \hline
    11 & 5.31222269340309e-26 & 1.63787475883274e-23 & 7.60709149843016e-27 & 1.48524338178853e-28 \\
    \hline
    12 & 0.497709058034396 & 1.6648036240588e-09 & 1.69282426111285e-22 & 1.33481951207468e-28 \\
    \hline
    13 & 0.566249724837391 & 0.000616314917991707 & 2.80559368851122e-08 & 3.73468872536468e-10 \\
    \hline
    14 & 4.01580837957093e-52 & 2.47104714917706e-39 & 4.489035680322e-30 & 4.49561835420922e-27 \\
    \hline
    15 & 4.56729725440379e-39 & 9.76965897576114e-41 & 1.21661809077255e-49 & 1.06476400936823e-56 \\
    \hline
    16 & 7.47000341630564e-31 & 1.58641432724581e-15 & 1.52696574288767e-05 & 0.11260123563263 \\
    \hline
    17 & 7.83899570450732e-18 & 1.19259522961946e-19 & 2.96030193616771e-23 & 5.78387135065131e-25 \\
    \hline
    18 & 3.94573584948892e-49 & 9.74229409485932e-46 & 9.6222357975944e-47 & 5.15567530467909e-48 \\
    \hline
    19 & 0.000140819159672948 & 0.992456346616752 & 0.997291516875785 & 0.93537161833889 \\
    \hline
    20 & 7.19059102218209e-24 & 1.34315347447786e-16 & 1.04717335990073e-11 & 2.22300748226144e-10 \\
    \hline
    \end{tabular}
\end{table}

\begin{table}[h]
    \centering
    \caption{Strain abundance profile for 2 hosts each with 10 strains, $t=1500$}
    \label{2-10-3}
    \begin{tabular}{|c|c|c|c|c|}
    \hline
    strain & $m=0.025$ & $m=0.05$ & $m=0.075$ & $m=0.1$ \\
    \hline
    1 & 4.19743032452484e-91 & 1.98484075372728e-92 & 4.32515626996602e-95 & 4.610440219831e-95 \\
    \hline
    2 & 1.67061522131889e-77 & 6.9335006584254e-74 & 4.57148033793554e-74 & 1.45102893189806e-73 \\
    \hline
    3 & 2.96760325240378e-36 & 5.812127147615e-33 & 1.1456547080284e-32 & 4.00538019719074e-32 \\
    \hline
    4 & 1.33477240435138e-58 & 1.7353203496624e-56 & 1.77610664226523e-56 & 4.19184801609013e-56 \\
    \hline
    5 & 6.96915947017103e-43 & 9.10616430137575e-53 & 3.58689150286478e-61 & 1.27960266961761e-66 \\
    \hline
    6 & 0.675803166194659 & 0.675753077885559 & 0.675783194621681 & 0.675593140071391 \\
    \hline
    7 & 3.74056481863755e-95 & 1.98438843131503e-94 & 9.58307467688429e-96 & 6.20064713725716e-95 \\
    \hline
    8 & 4.2538653510647e-74 & 1.73651809724369e-72 & 1.76596000115316e-72 & 3.64324173032911e-72 \\
    \hline
    9 & 0.589135629710945 & 0.589077735047019 & 0.589113010212312 & 0.588892806836195 \\
    \hline
    10 & 5.16394634737746e-93 & 2.34143879314971e-90 & 3.15913747927784e-89 & 9.72300057517985e-88 \\
    \hline
    11 & 6.79275551645223e-42 & 6.03224910571545e-47 & 6.03421674026248e-49 & 3.92435455348091e-49 \\
    \hline
    12 & 0.174495492764285 & 2.37851973269004e-40 & 1.79932426097318e-48 & 4.53493265362497e-42 \\
    \hline
    13 & 0.981642252262331 & 5.3251493463262e-20 & 6.54171560297791e-16 & 3.28899139755983e-06 \\
    \hline
    14 & 1.79975332250962e-85 & 2.69962283030498e-41 & 8.45988115543282e-45 & 1.27062270449315e-57 \\
    \hline
    15 & 6.63174547273489e-63 & 9.22209377009531e-54 & 8.22249892426513e-68 & 1.40290282717911e-73 \\
    \hline
    16 & 1.97993489627049e-57 & 5.0092368929387e-08 & 1.15161863795729 & 1.15222798851069 \\
    \hline
    17 & 7.89376854879037e-28 & 2.66784174315917e-43 & 1.43613983768721e-46 & 5.36460427237102e-48 \\
    \hline
    18 & 9.97412751355532e-79 & 7.98051633855547e-79 & 2.12884619135717e-68 & 4.80907454795742e-48 \\
    \hline
    19 & 1.14525706300827e-09 & 1.0563667656185 & 0.000540558090429951 & 9.61516965680835e-09 \\
    \hline
    20 & 2.46115038996684e-39 & 1.30933917419422e-15 & 4.37662237467484e-24 & 1.24170559850099e-40 \\
    \hline
    \end{tabular}
\end{table}

\section*{Acknowledgments}
L.S. would like to acknowledge the support of the Office of Research at Rutgers University-Camden.
\bibliographystyle{siamplain}
\bibliography{references}

\end{document}

%% file: ex_shared.tex
\usepackage{lipsum}
\usepackage{amsfonts}
\usepackage{graphicx}
\usepackage{epstopdf}
\usepackage{algorithmic}
\ifpdf
  \DeclareGraphicsExtensions{.eps,.pdf,.png,.jpg}
\else
  \DeclareGraphicsExtensions{.eps}
\fi

\usepackage{enumitem}
\setlist[enumerate]{leftmargin=.5in}
\setlist[itemize]{leftmargin=.5in}

\newsiamremark{remark}{Remark}
\newsiamremark{hypothesis}{Hypothesis}
\crefname{hypothesis}{Hypothesis}{Hypotheses}
\newsiamthm{claim}{Claim}
\newsiamremark{fact}{Fact}
\crefname{fact}{Fact}{Facts}

\headers{Modeling microbiome dynamics on social networks: how between-host transmission shapes within-host evolution}{Longmei Shu and Feng Fu}

\title{Modeling microbiome dynamics on social networks: how between-host transmission shapes within-host evolution\thanks{Submitted to the editors \today.
}}

\author{Longmei Shu\thanks{Department of Mathematical Sciences, Rutgers University, Camden, NJ
  (\email{longmei.shu@rutgers.edu}, \url{https://sites.rutgers.edu/longmei-shu/}).}
\and Feng Fu\thanks{Department of Mathematics \& Department of Biomedical Data Science, Dartmouth College, Hanover, NH
  (\email{Feng.Fu@dartmouth.edu}, \url{https://mhl.host.dartmouth.edu/index.html}).}
}

\usepackage{amsopn}

%% file: references.bib
@article{hmp2012,
  author  = {{Human Microbiome Project Consortium}},
  title   = {Structure, function and diversity of the healthy human microbiome},
  journal = {Nature},
  year    = {2012},
  volume  = {486},
  number  = {7402},
  pages   = {207--214},
  doi     = {10.1038/nature11234}
}

@article{yatsunenko2012,
  author  = {Yatsunenko, Tanya and Rey, Federico E. and Manary, Mark J. and Trehan, Indi and others},
  title   = {Human gut microbiome viewed across age and geography},
  journal = {Nature},
  year    = {2012},
  volume  = {486},
  number  = {7402},
  pages   = {222--227},
  doi     = {10.1038/nature11053}
}

@article{falony2016,
  author  = {Falony, Gwen and Joossens, Marie and Vieira-Silva, Sara and Wang, Jun and others},
  title   = {Population-level analysis of gut microbiome variation},
  journal = {Science},
  year    = {2016},
  volume  = {352},
  number  = {6285},
  pages   = {560--564},
  doi     = {10.1126/science.aad3503}
}

@article{zhernakova2016,
  author  = {Zhernakova, Alexandra and Kurilshikov, Alexander and Bonder, Marc Jan and Tigchelaar, Ettje F. and others},
  title   = {Population-based metagenomics analysis reveals markers for gut microbiome composition and diversity},
  journal = {Science},
  year    = {2016},
  volume  = {352},
  number  = {6285},
  pages   = {565--569},
  doi     = {10.1126/science.aad3369}
}

@article{faith2013,
  author  = {Faith, Jeremiah J. and Guruge, Janaki L. and Charbonneau, Mark and Subramanian, Sathish and Seedorf, Henning and Goodman, Andrew L. and Clemente, Jose C. and Knight, Rob and Heath, Andrew C. and Leibel, Rudolph L. and Rosenbaum, Michael and Gordon, Jeffrey I.},
  title   = {The long-term stability of the human gut microbiota},
  journal = {Science},
  year    = {2013},
  volume  = {341},
  number  = {6141},
  pages   = {1237439},
  doi     = {10.1126/science.1237439}
}

@article{dethlefsen2011,
  author  = {Dethlefsen, Les and Relman, David A.},
  title   = {Incomplete recovery and individualized responses of the human distal gut microbiota to repeated antibiotic perturbation},
  journal = {Proceedings of the National Academy of Sciences},
  year    = {2011},
  volume  = {108},
  number  = {Supplement 1},
  pages   = {4554--4561},
  doi     = {10.1073/pnas.1000087107}
}

@article{david2014,
  author  = {David, Lawrence A. and Maurice, Corinne F. and Carmody, Rachel N. and Gootenberg, David B. and Button, Julie E. and Wolfe, Benjamin E. and Ling, Alisha V. and Devlin, A. Sloan and Varma, Yug and Fischbach, Michael A. and Biddinger, Sudha B. and Dutton, Rachel J. and Turnbaugh, Peter J.},
  title   = {Diet rapidly and reproducibly alters the human gut microbiome},
  journal = {Nature},
  year    = {2014},
  volume  = {505},
  number  = {7484},
  pages   = {559--563},
  doi     = {10.1038/nature12820}
}

@article{vanNood2013,
  author  = {van Nood, Els and Vrieze, Anne and Nieuwdorp, Max and Fuentes, Susana and Zoetendal, Erwin G. and de Vos, Willem M. and Visser, Caroline E. and Kuijper, Ed J. and Bartelsman, Joep F. W. M. and Tijssen, Jan G. P. and Speelman, Peter and Dijkgraaf, Marcel G. W. and Keller, Josbert J.},
  title   = {Duodenal infusion of donor feces for recurrent {Clostridium difficile}},
  journal = {New England Journal of Medicine},
  year    = {2013},
  volume  = {368},
  number  = {5},
  pages   = {407--415},
  doi     = {10.1056/NEJMoa1205037}
}

@article{smillie2018,
  author  = {Smillie, Christopher S. and Sauk, Jenny and Gevers, Dirk and Friedman, Jonathan and Sung, Jaeyun and Youngster, Ilan and Hohmann, Elizabeth L. and Staley, Christopher and Khoruts, Alexander and Sadowsky, Michael J. and Allegretti, Jessica R. and Smith, Mark B. and Xavier, Ramnik J. and Alm, Eric J.},
  title   = {Strain Tracking Reveals the Determinants of Bacterial Engraftment in the Human Gut Following Fecal Microbiota Transplantation},
  journal = {Cell Host \& Microbe},
  year    = {2018},
  volume  = {23},
  number  = {2},
  pages   = {229--240.e5},
  doi     = {10.1016/j.chom.2018.01.003}
}

@article{song2013,
  author  = {Song, Se Jin and Lauber, Christian and Costello, Elizabeth K. and Lozupone, Catherine A. and Humphrey, Gregory and Berg-Lyons, Donna and Caporaso, J. Gregory and Knights, Dan and Clemente, Jose C. and Nakielny, Sara and Gordon, Jeffrey I. and Fierer, Noah and Knight, Rob},
  title   = {Cohabiting family members share microbiota with one another and with their dogs},
  journal = {eLife},
  year    = {2013},
  volume  = {2},
  pages   = {e00458},
  doi     = {10.7554/eLife.00458}
}

@article{tung2015,
  author  = {Tung, Jenny and Barreiro, Luis B. and Burns, Michael B. and Grenier, Jean-Christophe and Lynch, Josh and Grieneisen, Laura E. and Altmann, Jeanne and Alberts, Susan C. and Blekhman, Ran and Archie, Elizabeth A.},
  title   = {Social networks predict gut microbiome composition in wild baboons},
  journal = {eLife},
  year    = {2015},
  volume  = {4},
  pages   = {e05224},
  doi     = {10.7554/eLife.05224}
}

@article{moeller2016,
  author  = {Moeller, Andrew H. and Foerster, Steffen and Wilson, Michael L. and Pusey, Anne E. and Hahn, Beatrice H. and Ochman, Howard},
  title   = {Social behavior shapes the chimpanzee pan-microbiome},
  journal = {Science Advances},
  year    = {2016},
  volume  = {2},
  number  = {1},
  pages   = {e1500997},
  doi     = {10.1126/sciadv.1500997}
}

@article{brito2019,
  author  = {Brito, Ilana L. and Gurry, Thomas and Zhao, Shijie and Huang, Katherine and Young, Sarah K. and Shea, Terrence P. and Naisilisili, Waisea and Jenkins, Aaron P. and Jupiter, Stacy D. and Gevers, Dirk and Alm, Eric J.},
  title   = {Transmission of human-associated microbiota along family and social networks},
  journal = {Nature Microbiology},
  year    = {2019},
  volume  = {4},
  number  = {6},
  pages   = {964--971},
  doi     = {10.1038/s41564-019-0409-6}
}

@article{vallescolomer2023,
  author  = {Valles-Colomer, Mireia and Blanco-Míguez, Aitor and Manghi, Paolo and Asnicar, Francesco and others},
  title   = {The person-to-person transmission landscape of the gut and oral microbiomes},
  journal = {Nature},
  year    = {2023},
  volume  = {614},
  number  = {7946},
  pages   = {125--135},
  doi     = {10.1038/s41586-022-05620-1}
}

@article{coyte2015,
  author  = {Coyte, Katharine Z. and Schluter, Jonas and Foster, Kevin R.},
  title   = {The ecology of the microbiome: Networks, competition, and stability},
  journal = {Science},
  year    = {2015},
  volume  = {350},
  number  = {6261},
  pages   = {663--666},
  doi     = {10.1126/science.aad2602}
}

@article{stein2013,
  author  = {Stein, Richard R. and Bucci, Vanni and Toussaint, Nora C. and Buffie, Charlie G. and R{\"a}tsch, Gunnar and Pamer, Eric G. and Sander, Chris and Xavier, Jo{\~a}o B.},
  title   = {Ecological Modeling from Time-Series Inference: Insight into Dynamics and Stability of Intestinal Microbiota},
  journal = {PLoS Computational Biology},
  year    = {2013},
  volume  = {9},
  number  = {12},
  pages   = {e1003388},
  doi     = {10.1371/journal.pcbi.1003388}
}

@article{bucci2016,
  author  = {Bucci, Vanni and Tzen, Belinda and Li, Ning and Simmons, Matt and Tanoue, Takeshi and Bogart, Elijah and Deng, Luxue and Yeliseyev, Vladimir and Delaney, Mary L. and Liu, Qing and Olle, Bernat and Stein, Richard R. and Honda, Kenya and Bry, Lynn and Gerber, Georg K.},
  title   = {{MDSINE}: Microbial Dynamical Systems INference Engine for microbiome time-series analyses},
  journal = {Genome Biology},
  year    = {2016},
  volume  = {17},
  pages   = {121},
  doi     = {10.1186/s13059-016-0980-6}
}

@article{venturelli2018,
  author  = {Venturelli, Ophelia S. and Carr, Alex and Fisher, Garth and Hsu, Ryan H. and Lau, Rebecca and Bowen, Benjamin P. and Hromada, Susan and Northen, Trent and Arkin, Adam P.},
  title   = {Deciphering microbial interactions in synthetic human gut microbiome communities},
  journal = {Molecular Systems Biology},
  year    = {2018},
  volume  = {14},
  number  = {6},
  pages   = {e8157},
  doi     = {10.15252/msb.20178157}
}

@article{buffie2015,
  author  = {Buffie, Charlie G. and Bucci, Vanni and Stein, Richard R. and McKenney, Peter T. and Ling, Lilan and Gobourne, Asia and No, Daniel and Liu, Hui and Kinnebrew, Melissa and Viale, Agnes and Littmann, Eric and van den Brink, Marcel R. M. and Jenq, Robert R. and Taur, Ying and Sander, Chris and Cross, Justin R. and Toussaint, Nora C. and Xavier, Jo{\~a}o B. and Pamer, Eric G.},
  title   = {Precision microbiome reconstitution restores bile acid mediated resistance to {Clostridium difficile}},
  journal = {Nature},
  year    = {2015},
  volume  = {517},
  number  = {7533},
  pages   = {205--208},
  doi     = {10.1038/nature13828}
}

@article{bashan2016,
  author  = {Bashan, Amir and Gibson, Travis E. and Friedman, Jonathan and Carey, Vincent J. and Weiss, Scott T. and Hohmann, Elizabeth L. and Liu, Yang-Yu},
  title   = {Universality of human microbial dynamics},
  journal = {Nature},
  year    = {2016},
  volume  = {534},
  number  = {7606},
  pages   = {259--262},
  doi     = {10.1038/nature18301}
}

@article{coyte19,
  author  = {Coyte, Katharine Z. and Rakoff-Nahoum, Seth},
  title   = {Understanding Competition and Cooperation within the Mammalian Gut Microbiome},
  journal = {Current Biology},
  year    = {2019},
  volume  = {29},
  number  = {11},
  pages   = {R538--R544},
  doi     = {10.1016/j.cub.2019.04.017}
}

@article{figueiredo20,
  author  = {Figueiredo, Alexandre R. T. and Kramer, Jos},
  title   = {Cooperation and Conflict Within the Microbiota and Their Effects On Animal Hosts},
  journal = {Frontiers in Ecology and Evolution},
  year    = {2020},
  volume  = {8},
  pages   = {132},
  doi     = {10.3389/fevo.2020.00132}
}

@article{wang241,
  author  = {Wang, Jia and Appidi, Manasa R. and Burdick, Leah H. and Abraham, Paul E. and Hettich, Robert L. and Pelletier, Dale A. and Doktycz, Mitchel J.},
  title   = {Formation of a constructed microbial community in a nutrient-rich environment indicates bacterial interspecific competition},
  journal = {mSystems},
  year    = {2024},
  volume  = {9},
  number  = {4},
  doi     = {10.1128/msystems.00006-24}
}

@article{johnson25,
  author        = {Johnson, Michael and Porter, Mason A.},
  title         = {Interacting Hosts with Microbiome Exchange: An Extension of Metacommunity Theory for Discrete Interactions},
  year          = {2025},
  eprint        = {2507.11958},
  archivePrefix = {arXiv},
  primaryClass  = {math.DS},
  note          = {arXiv:2507.11958}
}

@article{beghini25,
    author = {Francesco Beghini and Jackson Pullman and Marcus Alexander and Shivkumar Vishnempet Shridhar and Drew Prinster and Adarsh Singh and Rigoberto Matute Ju\'arez and Edwardo M. Airoldi and Ilana L. Brito and Nicholas A. Christakis},
    title = {Gut microbiome strain-sharing within isolated village social networks},
    journal = {Nature},
    volume = {637},
    pages = {167-175},
    year = {2025}
}

@article{hofbauer22,
    author = {Josef Hofbauer and Sebastian J. Schreiber},
    title = {Permanence via invasion graphs: incorporating community assembly into modern coexistence theory},
    journal = {Journal of Mathematical Biology},
    year = {2022},
    volume = {85},
    pages = {54}
}
